\documentclass[prd,%
preprint,%
tightenlines,superscriptaddress,showpacs,%
nofootinbib,eqsecnum,amsfonts,amsmath]{revtex4}
\usepackage{bm}

\begin{document}
\title{The 2.5PN gravitational wave polarisations from\\ inspiralling
compact binaries in circular orbits}


\author{K G Arun} \email{arun@rri.res.in} \affiliation{Raman Research
Institute, Bangalore 560 080, India} \author{Luc Blanchet}
\email{blanchet@iap.fr} \affiliation{${\mathcal{G}}{\mathbb{R}}
\varepsilon{\mathbb{C}}{\mathcal{O}}$, Institut d'Astrophysique de
Paris --- C.N.R.S., 98$^{\text{bis}}$ boulevard Arago, 75014 Paris,
France} \author{Bala R Iyer} \email{bri@rri.res.in} \affiliation{Raman
Research Institute, Bangalore 560 080, India} \author{Moh'd S S
Qusailah}\altaffiliation{On leave from University of Sana, Yemen}
\email{mssq@rri.res.in} \affiliation{Raman Research Institute,
Bangalore 560 080, India}
\begin{abstract}
Using the multipolar post-Minkowskian and matching formalism we
compute the gravitational waveform of inspiralling compact binaries
moving in quasi-circular orbits at the second and a half
post-Newtonian (2.5PN) approximation to general relativity. The inputs
we use include notably the mass-type quadrupole at the 2.5PN order,
the mass octupole and current quadrupole at the 2PN order, the mass
$2^5$-pole and current $2^4$-pole at 1PN. The non-linear hereditary
terms come from the monopole-quadrupole multipole interactions or
tails, present at the 1.5PN, 2PN and 2.5PN orders, and the
quadrupole-quadrupole interaction arising at the 2.5PN level. In
particular, the specific effect of non-linear memory is computed using
a simplified model of binary evolution in the past. The ``plus'' and
``cross'' wave polarisations at the 2.5PN order are obtained in
ready-to-use form, extending the 2PN results calculated earlier by
Blanchet, Iyer, Will and Wiseman.
\end{abstract}
\pacs{04.25.Nx, 04.30.-w, 97.60.Jd, 97.60.Lf}
\preprint{gr-qc/0404085}
\maketitle

\widetext
\baselineskip22pt

\section{Introduction}\label{sec:intro}

\subsection{On gravitational-wave astronomy}\label{sec:astro}

With the ongoing scientific runs of the Laser Interferometric
Gravitational Wave Observatory (LIGO) \cite{ligo} and the recent
commissioning of the French-Italian VIRGO detector \cite{virgo}, there
is tremendous excitement among physicists and astronomers about the
possibility of the first ever direct detection of gravitational waves.
The Japanese detector TAMA300 is already taking data \cite{tama} and
the British-German joint venture GEO600 will be operational soon
\cite{geo}. These detectors will form a world wide network capable of
coincident observations and joint data analysis. The design
sensitivity of the three LIGO interferometers together with that of
VIRGO and GEO600 is sufficient for the detection of GWs from powerful
astrophysical sources like binary black holes in the mass range of
about 40 $M_{\odot}$ \cite{Grish,DIS3,BCV}. These detectors are ground based
and will be sensitive to the high frequency regime of the spectrum of
GW, {\it i.e.} between $10\,$Hz - $10\,$kHz. The proposed space-borne
detector LISA (Laser Interferometric Space Antenna) will probe the
universe at low frequencies ($1\,$Hz - $10^{-4}\,$Hz)
\cite{lisa}. Even if LIGO-1 will not detect any gravitational waves,
it will be able to put upper bounds on the fluxes and populations of
different types of sources. The recent discovery of a binary pulsar
PSR J0737-3039 \cite{newbinpulsar} has brought double neutron star
coalescence rate estimates within an astrophysically relevant regime
for LIGO-1 and VIRGO. The estimated event rate for compact binary
merger for LIGO-1 can now go as high as 1 event per 1.5yr
\cite{kalogera}. LIGO-2 with its ten fold improved sensitivity should
definitely be able to detect gravitational waves from different
astrophysical sources.

GW signals can be broadly classified into four categories: inspiral,
periodic, burst and stochastic. The strategies for data analysis vary
for different types of signals. Deterministic signals from inspiral
and periodic sources will be probed by matched filtering (to be
discussed later), whereas statistical signals such as stochastic
sources will be searched for by cross-correlating pairs of
interferometers, seeking correlations and statistical
variations. Unmodelled signals like those from supernova explosion or
gamma ray bursts (GRBs) cannot in principle be looked for by any
parametric techniques. One needs to monitor power excesses in the
frequency domain or have notable amplitude variations in time to
detect them. In all these studies, coincidence between multiple
detectors improves the efficiency of detection significantly.

The joint observation of astrophysical phenomena in electromagnetic as
well as non-electromagnetic windows (such as GWs and neutrinos) will
have a huge impact on observational astronomy in the future. The
current theories of GRB's, involve either the merger of a compact
binary or a collapse. In either case, the GRBs are associated with the
emission of gravitational radiation \cite{meszaros}. Detection of a
merger GW signal would require a lower signal-to-noise ratio, if it is
coincident with a GRB \cite{kochanek}. The data analysis techniques
for such phenomena will include comparison of the correlated output of
two gravitational wave detectors, one prior to the GRB and the other
at times not associated with the GRB \cite{grb2,grb3}. The spectral
flux distribution of the neutrinos from GRBs associated with
supernovae, can test for the possible time delays between the
supernovae and the GRB down to much shorter time scales than can be
resolved with only photons \cite{neutrino}. 

In the cosmological context, the primordial gravitational waves
resulting from inflation, contribute to the polarisation of the cosmic
microwave background radiation (CMBR). Their observations
could provide us valuable information about the early universe as well
as important checks on the inflationary cosmological model
\cite{Lewis,cmb1,cmb2}. The proposed Decihertz Interferometer Gravitational
Wave Observatory (DECIGO) \cite{accn}, could even probe the
acceleration of the universe, and hence provide an independent
determination of the spatial curvature of the cosmological model. This
is important as projects like SuperNova/Acceleration Probe (SNAP)
\cite{snap} would also probe the acceleration of the universe by
electromagnetic means. 

Pulsars can spin up by accreting matter from the neighbouring
star. Recent observations point towards a possibility of GW emission
balancing the accretion torque putting upper bounds on pulsar spin
\cite{pulsar}. If these gravitational waves are due to r-mode or CFS
instability in the pulsar, as it is presently understood, they can be
detected by the advanced LIGO detector \cite{And-Kokk}. The detection
of GWs from instabilities in relativistic stars \cite{Andersson} by
advanced LIGO would be able to test models of stellar interiors.

The ``chirp'' from the inspiral of two compact objects is one of the
most plausible GW signals the ground based detectors would look for.
The early inspiral will fall in the sensitivity band of the
space-based detectors, where as, the late inspiral will be a good
candidate source for the ground-based detectors. Though these GWs are
extremely weak and buried deep in the detector noise, the large number
of precisely predictable cycles in the detector bandwidth would push
the signal up to the level of detection. One can then use the
technique of matched filtering first for the detection of GW and later
({\it i.e.} off-line) for the estimation of the parameters of the
binary. In order to have a good detection, it is extremely important
to cross-correlate the detector output with a number of copies of the
theoretically predicted signal (corresponding to different signal
parameters) which is as precise as possible, and which remains in
accurate phase with the signal. This has made general relativistic
modelling of the inspiralling compact binary (ICB) one of the most
demanding requirements for GW data analysis
\cite{3mn,CF,P95,PW95,KKS95,DIS1,DIS2,DIS3}.

\subsection{An overview of current calculations of radiation from ICBs}
\label{sec:overview}

Currently, post-Newtonian (PN) theory provides the most satisfactory
description of the dynamics of ICBs and gravitational radiation
emitted by them. Starting from the gravitational-wave generation
formalism based on multipolar post-Minkowskian expansions (see the
next Section), the gravitational waveform and energy flux at the 2PN
order\,\footnote{As usual the $n$PN order refers either to the terms
$\sim 1/c^{2n}$ in the equations of motion, with respect to the usual
Newtonian acceleration, or in the radiation field, relatively to the
standard quadrupolar waveform.} were computed by Blanchet, Damour and
Iyer \cite{BDI}. This incorporated the tail contribution at 1.5PN
order both in the waveform and in the energy flux; the polarisation
states corresponding to the 1.5PN waveform were calculated in
Ref. \cite{Wiseman} (note that some algebraic errors in this reference
are corrected in \cite{BIWW}). The 2PN results have been independently
obtained using a direct integration of the relaxed Einstein field
equation \cite{WWWF,BDIWW}. The associated polarisation states ({\it
i.e.} the ``plus'' and ``cross'' polarisation waveforms) were obtained
in Ref. \cite{BIWW}. These works provided accurate theoretical
templates which are currently used for data analysis in all the laser interferometric GW detectors like LIGO/VIRGO.
Extending the wave-generation formalism, the 2.5PN term in
the energy flux, which arises from a subdominant tail effect, was
added in Ref. \cite{Beflux-96}. In the case of binaries moving in
quasi-elliptical orbits, the instantaneous parts of the waveform,
energy flux and angular momentum flux at 2PN order were computed by
Gopakumar and Iyer \cite{GI}. The polarisations of the waveform at
this order (in the adiabatic approximation) has been obtained more
recently \cite{GI-pol}, and the phasing of binaries in inspiralling
eccentric orbits has also been discussed \cite{DGI03}.

The extension of the gravitational wave generation formalism to third
post-Newtonian order, and computation of the energy flux up to 3.5PN
accuracy, was achieved in \cite{BIJ}. To this order, in addition to
the ``instantaneous'' contributions, coming from relativistic
corrections in the multipole moments of the source, the results
include several effects of tails, and tails generated by tails. But,
unlike at the 2PN or 2.5PN order, where the calculation is free from
ambiguities, at the 3PN order the incompleteness of the Hadamard
self-field regularisation leads to some undetermined constants in the
mass quadrupole moment of point particle binaries (we comment more on
this below). On the other hand, the computation of the binary's flux
crucially requires the 3PN equations of motion (EOM). These were
obtained earlier by two independent calculations, one based on the ADM
Hamiltonian formalism of general relativity
\cite{JS98,JS99,DJS00,DJS01}, the other on the direct 3PN iteration of
the Einstein field equations in harmonic coordinates
\cite{BF00,BFeom,ABF,BI-03}. Both approaches lead to an undetermined
constant parameter at 3PN when using a Hadamard regularisation, but
this constant has now been fixed using a dimensional regularisation
\cite{dimreg,BDE03}. An independent method \cite{IFA,itoh1,itoh2},
using surface integrals together with a strong field point particle
limit, has yielded results for the 3PN EOM in agreement with those of
the first two methods. In particular, the EOM so obtained are
independent of any ambiguity parameter, and consistent with the end
result of dimensional regularisation \cite{dimreg,BDE03}. The
conserved 3PN energy is thus uniquely determined, and then the 3.5PN
energy flux, together with the usual energy balance argument leads to
the expression to the evolution for orbital phase and frequency at
3.5PN order \cite{BFIJ}.

Once the gravitational wave polarisations are available, one proceeds to
use them to construct templates for gravitational-wave data analysis.
Templates for detection generally use the so-called restricted
waveform (RWF), where the phase is at the highest available PN order,
but the amplitude is retained to be Newtonian, involving only the main
signal harmonics at twice the orbital frequency. Though for detection
(which will be done online) the RWF approximation may be enough
\cite{CF,DIS3,BCV,DIJS}, the complete waveform will be useful for the
parameter estimation (off-line). Recently, studies using the complete
waveform, which includes the contribution from higher harmonics
besides the dominant one, have shown that it may play a vital role in
parameter estimation \cite{HM1,HM2,sintes}. The complete waveform
carries information which can break the degeneracy of the model, and
allow one to estimate the otherwise badly correlated parameters. In
the case of a chirping neutron star binary, the masses of the
individual stars can be separated because of the mass dependence of
the higher harmonics \cite{HM1}. In the case of black hole binaries,
whose frequencies are too low to be seen in the detector sensitivity
window for long, higher harmonics compensate for the information lost
when the signal does not last long enough to be apparent in the data
\cite{HM1}. An independent study \cite{HM2} about the angular
resolution of the space-based LISA-type gravitational wave detectors
with a time domain 2PN waveform, showed the importance of including
higher PN corrections to the wave amplitude in predicting the angular
resolution of the elliptic-plane detector configuration.

In the present work we provide the gravitational waveform from ICBs to
still higher accuracy, namely 2.5PN, which should in consequence be
useful for future improved studies in GW data analysis, for both
LIGO-type and LISA-type detectors. We shall include in the 2.5PN
waveform instantaneous as well as ``hereditary'' terms, exactly as
they are predicted by general relativity, completing therefore the
2.5PN generation problem for binaries moving in quasi-circular orbits
initiated in \cite{Beflux-96}. Using the waveform we next obtain the
two ``plus'' and ``cross'' GW polarisations at 2.5PN extending the
results of \cite{BIWW}. We shall verify that the 2.5PN wave form is in
perfect agreement, in the test-mass limit for one of the bodies, with the
result of linear black hole perturbations \cite{TS94}.

The plan of this paper is as follows. In Section \ref{sec:generation},
we outline the post-Newtonian generation formalism, based on
multipolar post-Minkowskian (MPM) expansions and matching to a general
post-Newtonian source. In Section \ref{sec:GWF}, we discuss the
structure of the gravitational waveform up to the 2.5PN order for
general sources. Then we present our list of results for the mass and
current-type multipole moments required for the computation of the
waveform in the case of point particle binaries. In Section
\ref{sec:hered}, we compute all the hereditary contributions relevant
to the 2.5PN waveform, which are made at this order of tails and the
non-linear memory integral. The results for the 2.5PN waveform are
given in Section \ref{sec:2.5PN pol}, ready to be implemented as 2.5PN
GW template in LIGO/VIRGO experiments. Section \ref{sec:2.5PN pol}
contains also some short discussion, concerning the multipole moments
and future issues related with the 3PN waveform.

\section{The post-Newtonian wave generation formalism}\label{sec:generation}

The wave generation formalism relates the gravitational waves observed
at a detector in the far-zone of the source to the stress-energy
tensor of the source. Successful wave-generation formalisms mix and
match approximation techniques from currently available
collections. These include post-Minkowskian (PM) methods,
post-Newtonian (PN) methods, multipole (M) expansions and
perturbations around curved backgrounds. A recent review
\cite{Blanchet-rev} discusses in detail the formalism we follow in the
computation of the gravitational field; we summarise below the main
features of this approach. This formalism has two independent aspects
addressing two different problems. The first aspect, is the general
method applicable to extended or fluid sources with compact support,
based on the mixed PM and multipole expansion (we call it a MPM
expansion), and matching to some PN source. The second aspect, is the
application to point particle binaries modelling ICBs.

\subsection{The MPM expansion and matching to a post-Newtonian 
source}\label{MPM}

To define the solution in the exterior of the source within the
complete non-linear theory we follow
Refs. \cite{BD86,B87,BD88,BD92,Bquad,Btail}, who built on earlier
seminal work of Bonnor \cite{WB} and Thorne \cite{Thorne-RMP}, to set
up the multipolar post-Minkowskian expansion. Starting from the
general solution to the linearized Einstein's equations in the form of
a multipolar expansion (valid in the external region), we perform a PM
iteration and treat individually each multipolar piece at any PM
order. In addition to terms evaluated at one retarded time, the
expression for the gravitational field also contains terms integrated
over the entire past ``history'' of the source or hereditary
terms. For the external field, the general method is not limited {\it
a priori} to PN sources. However, closed form expressions for the
multipole moments can presently only be obtained for PN sources,
because the exterior field may be connected to the inner field only if
there exists an ``overlapping'' region where both the MPM and PN
expansions are valid and can be matched together. For PN sources, this
region always exists and is the exterior ($r>a$) near ($r\ll\lambda$)
zone. After matching, it is found that the multipole moments have a
non-compact support owing to the gravitational field stress-energy
distributed everywhere up to spatial infinity. To include correctly
these contributions coming from infinity, the definition of the
multipole moments involves a finite part operation, based on analytic
continuation. This process is also equivalent to a Hadamard ``partie
finie'' of the integrals at the bound at infinity.

The formalism, notably the asymptotic matching procedure therein, has
been explored in detail and extended in a systematic way to higher PN
orders \cite{BD89,DI91a,Bgen,B98mult}. The final result of this
analysis is that, the physical post-Newtonian (slowly moving) source
is characterized by six symmetric and trace free (STF) time-varying
moments, denoted $\{I_L,\,J_L,\,W_L,\,X_L,\,Y_L,\,Z_L\}$,\footnote{As
usual $L=i_1i_2\cdots i_l$ denotes a multi-index made of $l$ spatial
indices (ranging from 1 to 3). The integer $l$ is referred to as the
multipolar order.} which are specified for each source in the form of
functionals of the formal PN expansion, up to any PN order, of the
stress-energy pseudo-tensor $\tau^{\mu\nu}$ of the material and
gravitational fields \cite{B98mult}. These moments parametrize the
linear approximation to the vacuum metric outside the source, which is
the first approximation in the MPM algorithm. In the linearized
gravity case $\tau^{\mu\nu}$ reduces to the compact-support matter
stress-energy tensor $T^{\mu\nu}$ and the expressions match perfectly
with those derived in Ref. \cite{DI-91}.

Starting from the complete set of six STF {\it source moments}
$\{I_L,\,J_L,\,W_L,\,X_L,\,Y_L,\,Z_L\}$, for which general expressions
can be given valid to any PN order, one can define a different set of
only two ``{\it canonical}'' source moments, denoted $\{M_L,\,S_L\}$,
such that the two sets of moments $\{I_L,\,\cdots,\,Z_L\}$ and
$\{M_L,\,S_L\}$ are physically equivalent. By this we mean that they
describe the same physical source, {\it i.e.} the two metrics,
constructed respectively out of $\{I_L,\,\cdots,\,Z_L\}$ and
$\{M_L,\,S_L\}$, differ by a mere coordinate transformation (are
isometric). However, the six general source moments
$\{I_L,\,\cdots,\,Z_L\}$ are closer rooted to the source because we
know their expressions as integrals over $\tau^{\mu\nu}$. On the other
hand, the canonical source moments $\{M_L,\,S_L\}$ are also necessary
because their use simplifies the calculation of the external
non-linearities. In addition, their existence shows that any radiating
isolated source is characterized by two and only two sets of
time-varying multipole moments \cite{Thorne-RMP,BD86}.

The MPM formalism is valid all over the weak field region outside the
source including the wave zone (up to future null infinity). It is
defined in harmonic coordinates. The far zone expansion at Minkowskian
future null infinity contains logarithms in the distance which are
artefacts of the harmonic coordinates. One can define, step by step in
the PM expansion, some {\it radiative} coordinates by a coordinate
transformation so that the log-terms are eliminated \cite{B87} and one
recovers the standard (Bondi-type) radiative form of the metric, from
which the {\it radiative moments}, denoted $\{U_L,\,V_L\}$, can be
extracted in the usual way \cite{Thorne-RMP}. The wave generation
formalism resulting from the exterior MPM field and matching to the PN
source is able to take into account, in principle, any PN correction
in both the source and radiative multipole moments. Nonlinearities in
the external field are computed by a post-Minkowskian algorithm. This
allows one to obtain the radiative multipole moments $\{U_L,\,V_L\}$,
as some non-linear functional of the canonical moments
$\{M_L,\,S_L\}$, and then of the actual source moments
$\{I_L,\,\cdots,\,Z_L\}$. These relations between radiative and source
moments include many non-linear multipole interactions as the source
moments mix with each other as the waves propagate from the source to
the detector. The dominant non-linear effect is due to the tails of
wave, made of coupling between non-static moments and the total mass
of the source, occurring at 1.5PN order ($\sim 1/c^3$) relative to the
leading quadrupole radiation \cite{BD92}. There is a corresponding
tail effect in the equations of motion of the source, occuring at
1.5PN order relative to the leading 2.5PN radiation reaction, hence at
4PN order ($\sim 1/c^8$) beyond the Newtonian acceleration
\cite{BD88}. At higher PN orders, there are different types of
non-linear multipole interactions, that are responsible for the
presence of some important hereditary ({\it i.e.} past-history
dependent) contributions to the waveform and energy flux.

A different wave-generation formalism from isolated sources, based on
direct retarded integration of Einstein's equations in harmonic
coordinates, is due to Will and Wiseman \cite{WWWF}, and provided
major improvement and elucidation of earlier investigations in the
same line \cite{EW,Thorne-RMP}. This formalism is based on different
source multipole moments (defined by integrals extending over the near
zone only), together with a different scheme for computing the
non-linearities in the external field. It has currently been completed
up to the 2PN order. At the most general level, {\it i.e.} for any PN
extended source and in principle at any PN order, the Will-Wiseman
formalism is completely equivalent to the present formalism based on
MPM expansions with asymptotic matching (see Section 5.3 in
\cite{Blanchet-rev} for the proof).

\subsection{Application to compact binary systems}\label{appl}

This application represents the second aspect of our approach. To this
end, in the first instance, the compact objects (neutron stars or
black holes) are modelled as point particles represented by Dirac
$\delta$-functions. Indeed one can argue that for compact objects the
effects of finite size and quadrupole distortion induced by tidal
interactions are higher order in the PN approximation. However, the
general formalism outlined in Section \ref{MPM}, is set up for a
continuous (smooth) matter distribution, with continuous $T^{\mu\nu}$,
and cannot be directly applied to point particles, since they lead to
divergent integrals at the location of the particles, when
$T^{\mu\nu}_{\rm point-particle}$ is substituted into the source
moments $\{I_L,\,\cdots,\,Z_L\}$. The calculation needs to be
supplemented by a prescription for removing the infinite part of the
integrals. Hadamard regularisation, based on Hadamard's notion of
partie finie, is what we employ. This is our ansatz for applying a
well-defined general ``fluid'' formalism to an initially ill-defined
point-particle source.

To summarise: A systematic analytical approximation scheme has been
set up for the calculation of waveforms and associated quantities from
point particles to the PN order required (or permitted by given
resources). A technical cost is the need to handle $\delta$-functions
in a non-linear theory, which is dealt with the Hadamard
regularisation scheme or a variant of it. However, we already
mentioned that at the 3PN order, subtleties arise due notably to the
so called non-distributivity of the Hadamard partie finie, which
resulted, as shown in \cite{BIJ}, in some ``ambiguities'' when
computing the 3PN mass-type quadrupole moment, which could entirely be
encoded into three undetermined numerical coefficients $\xi$,
$\kappa$, $\zeta$. These combined into the unique quantity
$\theta=\xi+2\kappa+\zeta$ in the 3PN energy flux for circular orbits.

The latter ambiguities are the analogues of the undetermined
parameters found in the binary's EOM at 3PN order, namely $\omega_k$
and $\omega_s$ in the canonical ADM approach \cite{JS98,JS99}, and
$\lambda$ in the harmonic-coordinates formalism \cite{BF00,BFeom}. The
parameter $\lambda$ is related to the ``static'' ambiguity $\omega_s$
by $\lambda = -\frac{3}{11}\omega_s - \frac{1987}{3080}$, while the
``kinetic'' ambiguity $\omega_k$ has been determined \cite{BF00} to
the value $\frac{41}{24}$ (see \cite{DJS01,ABF} for details). The
presence of the static ambiguity $\omega_s$ or, equivalently,
$\lambda$, is a consequence of the Hadamard regularisation scheme
which happens to become physically incomplete at the 3PN
order. Recently, Damour, Jaranowski and Sch\"afer \cite{dimreg}
proposed to use a better regularisation: {\it dimensional}
regularisation. This led them to a unique determination of $\omega_s$,
namely $\omega_s=0$. More recently \cite{BDE03}, the application of
dimensional regularisation to the computation of the EOM in harmonic
coordinates has led to the equivalent result for $\lambda$, which is
$\lambda=-\frac{1987}{3080}$. The EOM are thus completely determined
to the 3PN order within both the ADM and harmonic-coordinates
approaches using Hadamard regularisation supplemented by a crucial
argument from dimensional regularisation in order to fix the last
parameter.\footnote{Note that both calculations \cite{dimreg,BDE03}
are performed in the limit $\varepsilon\rightarrow 0$, where the
dimension of space is $d=3+\varepsilon$. The principle is to add to
the end results given by the Hadamard regularisation
\cite{JS98,JS99,BF00,BFeom} the {\it difference} between the
dimensional and Hadamard regularisations, which is specifically due to
the poles $\sim 1/\varepsilon$ and their associated finite part.} All
the 3PN conserved quantities are determined in
Refs. \cite{DJS01,ABF}. A 3PN accurate center-of-mass has been
constructed and used to reduce the conserved energy and angular
momentum \cite{BI-03}.

The numerical values of the radiation-field-related ambiguity coefficients
$\xi$, $\kappa$ and $\zeta$ introduced in \cite{BIJ}
 have also been determined \cite{BDEI04} using dimensional regularization, so that the PN corrections to phasing are completely determined to 3.5PN accuracy. However, as we shall work in the present paper on
the 2.5PN waveform, {\it i.e.} one half order before 3PN, these
ambiguities will not concern us here, and we shall find that no
ambiguity shows up in any of our calculations based on Hadamard's
regularisation. In fact, it can be shown that up to the 2.5PN order
Hadamard's regularisation as we shall employ here gives the same
result as would dimensional regularisation. The reason is that, in the
source multipole moments up to this order, there are no logarithmic
divergences occuring at the particles' locations, which correspond in
$d$ dimensions to poles in $\varepsilon =d-3$.

In the present work we will require, for the computation of the
time-derivatives of multipole moments, the EOM for the case of
circular orbits to 2.5PN accuracy. We denote the PN parameter in
harmonic coordinates by

\begin{equation}
\gamma\equiv\frac{Gm}{rc^2}\,,
\label{gam}\end{equation}
where $r=|{\bf x}|$ is the binary's separation (${\bf x}\equiv {\bf
y}_1-{\bf y}_2$ is the vectorial separation between the particles),
and $m=m_1+m_2$ is the total mass of the binary system. We pose
$X_1=\frac{m_1}{m}$, $X_2=\frac{m_2}{m}$, and
$\nu=X_1X_2$. Occasionally we also employ $\delta m=m_1-m_2$ so that
$\frac{\delta m}{m}=X_1-X_2$. Then the 2.5PN binary's acceleration
reads (${\bf v}\equiv {\bf v}_1-{\bf v}_2$)\,\footnote{We
systematically use the shorthand ${\cal O}(n)$ to mean a small
post-Newtonian remainder term of the order of ${\cal O}(c^{-n})$.}

\begin{eqnarray}\label{eq:EOM 2.5PN}
{d{\bf v}\over dt} = -\omega^2 \, {\bf x}-{32\,G^3\over
5\,c^5}\,\frac{m^3\,\nu}{r^4}\,{\bf v} + {\cal O}(6)\,,
\end{eqnarray}
where the explicit 2.5PN term ($\sim 1/c^5$) is the radiation reaction
force in the harmonic coordinate system used here. The radiation
reaction force plays an important role in our calculation of the
waveform --- it must be consistently included in all replacements of
accelerations at 2.5PN order (however the reaction force yields no
contribution to the energy flux at 2.5PN order for circular orbits
\cite{Beflux-96}). In Eq.~(\ref{eq:EOM 2.5PN}) the orbital frequency
$\omega\equiv 2\pi/P$ (where $P$ is the orbital period) is related to
the binary's separation $r$ in harmonic coordinates with 2PN accuracy
by \cite{DD}

\begin{equation}\label{eq:2pn-omega}
\omega^2 = {G\,m\over r^3} \left\{1+\Bigl[-3+\nu\Bigr]\gamma +
\biggl[6+{41\over 4} \nu + \nu^2 \biggr] \gamma^2 +{\cal O}(6)\right\}
\,.
\end{equation}
In the following we shall also need the inverse of
Eq.~(\ref{eq:2pn-omega}), {\it i.e.} $\gamma$ in terms of $\omega$,
which can conveniently be written into the form
\begin{equation}\label{eq:gamma-x}
\gamma = x \left\{1+\Bigl[1-{\nu \over3}\Bigr] x + \biggl[1-{65\over
12}\nu \biggr] x^2 +{\cal O}(6)\right\} \,,
\end{equation}
in which we have introduced the gauge invariant frequency-dependent
parameter

\begin{equation}\label{x}
x\equiv \left({G\,m\,\omega \over c^3}\right)^{2/3}\,.
\end{equation}

\section{The 2.5PN gravitational waveform}\label{sec:GWF}

\subsection{Waveform as a functional of multipole moments}\label{WFmult}

In an appropriate radiative coordinate system $X^\mu=(c T,X^i)$, the
transverse-traceless (TT) projection of the deviation of the metric of
an isolated body from flat metric defines the asymptotic waveform
$h_{km}^{\rm TT}$ (lower-case Latin indices take the values
$1,2,3$). The leading-order $1/R$ part of $h_{km}^{\rm TT}$ (where
$R=|\mathbf{X}|$ is the distance to the body) can be uniquely
decomposed \cite{Thorne-RMP} into its {\it radiative} multipole
contributions introduced in Section \ref{sec:generation}. Furthermore,
the PN order of the asymptotic waveform scales with the multipolar
order $l$. Hence, at any PN order only a finite number of multipoles
is required, and we have, with 2.5PN accuracy,

\begin{eqnarray}\label{eq:WFgen}
h^{\rm TT}_{km} &=& {2G\over c^4R} \,{\cal P}_{ijkm}({\bf N}) \biggl\{
U_{ij} \nonumber \\ &&\quad\qquad + {1\over c} \left[ {1\over 3} N_a
U_{ija} + {4\over 3} \varepsilon_{ab(i} V_{j)a} N_b \right] \nonumber
\\ &&\quad\qquad + {1\over c^2} \left[ {1\over 12} N_{ab} U_{ijab} +
{1\over 2} \varepsilon_{ab(i} V_{j)ac} N_{bc} \right] \nonumber \\
&&\quad\qquad + {1\over c^3} \left[ {1\over 60} N_{abc} U_{ijabc} +
{2\over 15} \varepsilon_{ab(i} V_{j)acd} N_{bcd} \right] \nonumber \\
&&\quad\qquad + {1\over c^4} \left[ {1\over 360} N_{abcd} U_{ijabcd} +
{1\over 36} \varepsilon_{ab(i} V_{j)acde} N_{bcde} \right]\nonumber \\
&&\quad\qquad +{1\over c^5} \left[ {1\over 2520} N_{abcde} U_{ijabcde}
+ {1\over 210} \varepsilon_{ab(i} V_{j)acdef} N_{bcdef} \right] +
{\cal O}(6) \biggr\} \,.
\end{eqnarray}
The $U_L$'s and $V_L$'s (with $L=ij\cdots$ a multi-index composed of
$l$ indices) appearing in the above waveform are respectively called
the mass-type and the current-type radiative multipole moments (see
discussion in Section \ref{MPM}). They are functions of the retarded
time $T_R\equiv T-R/c$ in radiative coordinates, $U_L(T_R)$ and
$V_L(T_R)$. We denote by ${\bf N}\equiv {\bf X}/R$ the unit vector
pointing along the direction of the source located at distance $R$
from the detector. A product of components of ${\bf
N}=(N_i)_{i=1,2,3}$ is generally denoted $N_L\equiv N_iN_j\cdots$. The
Levi-Civita antisymmetric symbol reads $\varepsilon_{abi}$, such that
$\varepsilon_{123}=+1$. The operator ${\cal P}_{ijkm}$ represents the
usual TT algebraic projector which is given by

\begin{subequations}\label{Pijkl}\begin{eqnarray}
{\cal P}_{ijkm} &=& {1\over 2}\bigl({\cal P}_{ik}{\cal P}_{jm}+{\cal
P}_{im}{\cal P}_{jk}-{\cal P}_{ij}{\cal P}_{km}\bigr)\,,\\ {\cal
P}_{ij} &\equiv& \delta_{ij} -N_iN_j\,.
\end{eqnarray}\end{subequations}
Given an orthonormal triad $({\bf N},\,{\bf P},\,{\bf Q})$, consisting
of the radial direction ${\bf N}$ to the observer, and two unit
polarisation vectors ${\bf P}$ and ${\bf Q}$, transverse to the
direction of propagation, we define the two ``plus'' and ``cross''
polarisation waveforms by

\begin{subequations}\label{plus-cross}
\begin{eqnarray}
h_+ = {1 \over 2} (P_i P_j - Q_i Q_j) h_{ij}^{\rm TT} \,, \\ h_\times
= {1 \over 2} (P_i Q_j + Q_i P_j) h_{ij}^{\rm TT} \,,
\end{eqnarray}
\end{subequations}
in which the projector ${\cal P}_{ijkm}$ present in front of the TT
waveform may be omitted. 

In the case of circular binary systems we shall adopt for ${\bf P}$
the vector lying along the intersection of the orbital plane with the
plane of the sky in the direction of the ``ascending node'' ${\cal
N}$, {\it i.e.} the point at which the bodies cross the plane of the
sky moving toward the detector, and ${\bf Q} = {\bf N} \times {\bf
P}$. Following the convention of Ref. \cite{BIWW}, the unit vector
joining the particle 2 to the particle 1, {\it i.e.} ${\bf n}=({\bf
y}_1-{\bf y}_2)/r$ where $r=|{\bf y}_1-{\bf y}_2|$, is given
by ${\bf n}= {\bf P} \cos\phi + ({\bf Q} \cos i + {\bf N} \sin i) \sin
\phi$, where $i$ denotes the orbit's inclination angle and $\phi$ is
the orbital phase, namely the angle between the ascending node and the
direction of body one.\footnote{The angle $\phi$ in our convention
differs by ${\pi \over2}$ from the same in Refs. \cite{BDI,GI-pol}. We
follow here the convention of BIWW \cite{BIWW}, that is related to the
BDI one \cite{BDI,GI-pol} by $\phi_{\rm BDI} = \phi_{\rm BIWW}
-{\pi\over 2}$.} The unit direction of the velocity, {\it i.e.}
$\bm{\lambda}$ such that ${\bf v}=r\omega \bm{\lambda}$ (for circular
orbits), is given by $\bm{\lambda}= -{\bf P} \sin \phi + ({\bf Q} \cos
i + {\bf N} \sin i) \cos \phi$. (See Fig. 7 and Eq. (7.4) of
\cite{WWWF}.)

Using the MPM formalism, the radiative moments entering
Eq.~(\ref{eq:WFgen}) can be expressed in terms of the source variables
with sufficient accuracy, that is a fractional accuracy of ${\cal
O}(6)\equiv {\cal O}(c^{-6})$ relative to the lowest-order quadrupolar
waveform. For this approximation to be complete, one must compute:
mass-type radiative quadrupole $U_{ij}$ with 2.5PN accuracy;
current-type radiative quadrupole $V_{ij}$ and mass-type radiative
octupole $U_{ijk}$ with 2PN accuracy; mass-type hexadecapole
$U_{ijkl}$ and current-type octupole $V_{ijk}$ with 1.5PN precision;
$U_{ijklm}$ and $V_{ijkl}$ up to 1PN order; $U_{ijklmn}$, $V_{ijklm}$
at 0.5PN; and finally $U_{ijklmno}$, $V_{ijklmn}$ to Newtonian
precision. The relations connecting the radiative moments $U_L$ and
$V_L$ to the corresponding ``canonical'' moments $M_L$ and $S_L$ (see
Section \ref{MPM} for a short recall of their meaning) are given as
follows \cite{BD92,Bquad,Btail}. For the mass-type moments we have
{\allowdisplaybreaks
\begin{subequations}\label{eq:U}
\begin{eqnarray}
U_{ij}(T_R) &=& M^{(2)}_{ij} (T_R) + {2GM\over c^3}
\int_{-\infty}^{T_R} d V \left[ \ln \left({T_R-V\over
2b}\right)+{11\over12} \right] M^{(4)}_{ij} (V) \nonumber \\
&&+\frac{G}{c^5}\left\{-\frac{2}{7}\int_{-\infty}^{T_R} dV
M^{(3)}_{a<i}(V)M^{(3)}_{j>a}(V) \right.\nonumber \\ &&\qquad~\left.
+ {1 \over7}M^{(5)}_{a<i}M_{j>a} - {5 \over7}
M^{(4)}_{a<i}M^{(1)}_{j>a} -{2 \over7} M^{(3)}_{a<i}M^{(2)}_{j>a} +{1
\over3}\varepsilon_{ab<i}M^{(4)}_{j>a}S_{b}\right\}\nonumber \\ &&+
{\cal O}(6) \,,\label{eq:Uij}\\ U_{ijk} (T_R) &=& M^{(3)}_{ijk} (T_R)
+ {2GM\over c^3} \int_{-\infty}^{T_R} dV\left[ \ln \left({T_R-V\over
2b}\right)+{97\over60} \right] M^{(5)}_{ijk} (V)
\label{eq:Uijk}\nonumber \\ && +{\rm {\cal O}(5)} \,, \\ U_{ijkm}
(T_R) &=& M^{(4)}_{ijkm} (T_R) + {G\over c^3} \left\{ 2 M
\int_{-\infty}^{T_R} d V \left[ \ln \left({T_R-V\over
2b}\right)+{59\over30} \right] M^{(6)}_{ijkm}(V) \right.\nonumber\\
&&\quad\left. +{2\over5}\int_{-\infty}^{T_R} dV
M^{(3)}_{<ij}(V)M^{(3)}_{km>}(V)\right.\nonumber\\ &&\quad\left. -{21
\over5}M^{(5)}_{<ij}M_{km>}-{63 \over5}M^{(4)}_{<ij}M^{(1)}_{km>}-{102
\over5}M^{(3)}_{<ij}M^{(2)}_{km>}\right\}\nonumber \\ &&+ {\cal O}(4)
\,,
\end{eqnarray}\end{subequations}}\noindent
where the brackets $<>$ denote the symmetric-trace-free (STF)
projection, while, for the necessary current-type moments,
{\allowdisplaybreaks
\begin{subequations}\label{eq:V}
\begin{eqnarray}
V_{ij} (T_R) &=& S^{(2)}_{ij} (T_R) + {2GM\over c^3}
\int_{-\infty}^{T_R} d V \left[ \ln \left({T_R-V\over
2b}\right)+{7\over6} \right] S^{(4)}_{ij} (V)\label{eq:Vij}\nonumber
\\ &&+ {\cal O}(5) \,,\\
V_{ijk} (T_R) &=& S^{(3)}_{ijk} (T_R) +
{G\over c^3} \left\{ 2 M \int_{-\infty}^{T_R} d V \left[ \ln
\left({T_R-V\over 2b}\right)+{5\over3} \right] S^{(5)}_{ijk} (V)
\right.\nonumber \\ &&\quad\left.+{1\over
10}\varepsilon_{ab<i}M^{(5)}_{j\underline{a}}M_{k>b}-{1\over
2}\varepsilon_{ab<i}M^{(4)}_{j\underline{a}}M^{(1)}_{k>b} - 2
S_{<i}M^{(4)}_{jk>} \right\} \nonumber\\&& + {\cal O}(4) \,.
\end{eqnarray}\end{subequations}}\noindent
[The underlined index $\underline{a}$ means that it should be excluded
from the STF projection.] For all the other needed moments we are
allowed to simply write
\begin{subequations}\label{eq:UVl}
\begin{eqnarray}
U_L (T_R) &=& M^{(l)}_L(T_R) + {\cal O}(3)\,,\\ V_L (T_R) &=&
S^{(l)}_L(T_R) + {\cal O}(3)\,.
\end{eqnarray}
\end{subequations}
In the above formulas, $M$ is the total ADM mass of the binary system,
which agrees with the mass monopole moment. The $M_{L}$'s and
$S_L$'s are the mass and current-type canonical source moments, and
$M_{L}^{(p)}$, $S_{L}^{(p)}$ denote their $p$-th time derivatives.

The parameter $b$ appearing in the logarithms of Eqs.~(\ref{eq:U})
and (\ref{eq:V}) is a freely specifiable constant, having the
dimension of time, entering the relation between the retarded time
$T_R=T-R/c$ in radiative coordinates and the corresponding one
$t-\rho/c$ in harmonic coordinates (where $\rho$ is the distance of
the source in harmonic coordinates). More precisely we have
\begin{equation}
T_R = t- \frac{\rho}{c}
-\frac{2\,G\,M}{c^3}\ln\left(\frac{\rho}{c\,b}\right)\,.
\label{b}\end{equation}
The constant $b$ can be chosen at will because it simply corresponds
to a choice of the origin of radiative time with respect to harmonic
time.

As recalled in Section \ref{MPM}, the ``canonical'' moments $M_{L}$,
$S_L$ do not represent the best definition for what should be referred
to as the ``source'' moments. This is why we now replace the $M_{L}$'s
and $S_L$'s by adequate source multipole moments $I_{L}$, $J_L$,
$W_L$, $\cdots$, which admit closed-form expressions in terms of the
source's stress-energy tensor. At the 2.5PN order it turns out that we
need only to take into account a correction in the 2.5PN canonical
mass-type quadrupole moment and given, in a center-of-mass frame, by
(see Refs. \cite{Beflux-96,BIJ})\,\footnote{The equation (11.7a) in
\cite{BIJ} contains a sign error, but with no consequence for any of
the results of that reference. The correct sign is reproduced here.}
\begin{equation}
M_{ij}= I_{ij}+\frac{4G}{c^5}
\left[W^{(2)}I_{ij}-W^{(1)}I_{ij}^{(1)}\right] + {\cal O}(7),
\label{MijIij}
\end{equation}
where $I_{ij}$ is the source mass quadrupole, and where $W$ denotes
the ``monopole'' corresponding to the set of moments $W_L$. [We shall
need $W$ only at the Newtonian order where it will be given by
(\ref{monopoleW}); see Section \ref{sec:multipole moments} for the
expressions of all the source moments in the case of circular binary
systems.] Note that a formula generalizing Eq.~(\ref{MijIij}) to all
PN orders (and multipole interactions) is not possible at present and
needs to be investigated anew for specific cases. This is why it is
more convenient to define the source moments to be $I_L$ and $J_L$
(and the other ones $W_L,\,\cdots,\,Z_L$ as well, but in view of {\it
e.g.}  Eq.~(\ref{MijIij}) these appear to be much less important than
$I_L,\,J_L$) rather than $M_L$ and $S_L$. For all the other
moments needed here, besides the mass quadrupole (\ref{MijIij}), we can write,
with the required precision, that $M_L$ agrees with the corresponding
$I_L$ and that similarly $S_L$ agrees with $J_L$. Namely we always
have
\begin{subequations}
\label{MLIL}
\begin{eqnarray}
M_L &=& I_L + {\cal O}(5)\,,\\ S_L &=& J_L + {\cal O}(5)\,,
\end{eqnarray}
\end{subequations}
and we can neglect in the 2.5PN waveform all the remainders in
(\ref{MLIL}) except for the case of $M_{ij}$ where the required result
is provided by Eq.~(\ref{MijIij}). Thus, from now on, the waveform
will be considered a function of the ``main'' source moments $I_L$,
$J_L$, and also, of the ``auxiliary'' moment $W$ appearing in
Eq.~(\ref{MijIij}).

\subsection{Instantaneous versus hereditary contributions}
\label{insthered}

From Eqs.~(\ref{eq:U})--(\ref{eq:V}) it is clear that the radiative
moments contain two types of terms, those which depend on the source
moments at a single instant, namely the retarded time $T_{R}\equiv
T-R/c$, referred to as {\it instantaneous} terms, and the other ones
which are sensitive to the entire ``past history'' of the system, {\it
i.e.} which depend on all previous times ($V < T_R$), and are referred
to as the {\it hereditary} terms.

In this work, we find it convenient to further subclassify the
instantaneous terms in the radiative moments into three types based on
their structure. The leading instantaneous contribution to the
radiative moment $U_L$ (resp. $V_L$) from the source moment $I_L$
(resp. $J_L$), is of the form $I_L^{(l)}$ or $J_L^{(l)}$. We refer to
these as instantaneous contributions from the {\it source} moments and
denote them by the subscript ``${\rm inst(s)}$''. Starting at 2.5PN
order, additional instantaneous terms arise, of the form
$I_{ij}^{(n)}I_{km}^{(p)}$ or $I_{ij}^{(n)}J_k$, in the expressions
relating radiative moments to source moments [see
Eqs.~(\ref{eq:U})--(\ref{eq:V})]. We call such additional terms, the
instantaneous terms in the {\it radiative} moment and denote them by
the subscript ``${\rm inst(r)}$''. Thirdly, from Eq.~(\ref{MijIij}) we
see that the replacement of the canonical moments by the source
moments induces also some new terms at 2.5PN level, of the form
$I_{ij}^{(n)}W^{(p)}$, we shall call the instantaneous terms in the
{\it canonical} moment denoted by the subscript ``${\rm inst(c)}$''.

At the 1.5PN order, the hereditary terms are due to the interaction of
the mass quadrupole moment with mass monopole (ADM mass $M$) and
cause the effect of wave tails \cite{BD92}. Physically, this effect
can be visualized as the backscattering of the linear waves (described
by $I_{ij}$) off the constant spacetime curvature generated by the
mass energy $M$. This can be viewed as a part of the gravitational
field propagating inside the light cone ({\it e.g.} \cite{BD88}). At
higher PN orders there are similar tails due to the interaction between
$M$ and higher moments $I_{ijk}$, $J_{ij}$, $\cdots$. In addition, at
the 3PN order (however negligible for the present study), there is an
effect of tails generated by tails, because of the cubic interaction
between the quadrupole moment and two mass monopoles, $M\times M\times
I_{ij}$ \cite{Btail}. The hereditary term arising at the 2.5PN order
in the radiative quadrupole (\ref{eq:Uij}) is different in nature. It
is made of the quadrupole-quadrupole interaction, $I_{ij}\times
I_{kl}$, and can physically be thought of as due to the re-radiation
of the stress-energy tensor of the linearized quadrupolar
gravitational waves. It is responsible for the so-called ``non-linear
memory'' or Christodoulou effect \cite{chris,WW-chris,Thorne-chris}
(investigated within the present approach in \cite{BD92,Bquad}). So
far, all these effects are taken into account in the calculation of
the waveform up to 2PN order \cite{BDI} and in the energy flux up to
3.5PN \cite{Beflux-96,BIJ,BFIJ}. The two different types of hereditary
terms will be denoted by subscripts ``${\rm tail}$'' and ``${\rm
memory}$''.

Summarizing, with the present notation, the total 2.5PN waveform may
be written as
\begin{eqnarray}\label{eq:wf-parts}
h^{\rm TT}_{km} &=& \bigl(h^{\rm TT}_{km}\bigr)_{\rm inst(s)}+
\bigl(h_{km}^{\rm TT}\bigr)_{\rm inst(r)}+ \bigl(h_{km}^{\rm
TT}\bigr)_{\rm inst(c)}+\bigl(h_{km}^{\rm TT}\bigr)_{\rm tail}+
\bigl(h_{km}^{\rm TT}\bigr)_{\rm memory}\nonumber\\&+&{\cal O}(6) \,.
\end{eqnarray}
We give each of the above contributions explicitly. The instantaneous
part of type (s) reads
\begin{eqnarray}\label{eq:WFinst}
\bigl({ h^{\rm TT}_{km}}\bigr)_{\rm inst(s)} &=& {2G\over c^4R}
\,{\cal P}_{ijkm} \biggl\{ I^{(2)}_{ij} \nonumber \\ &&\qquad +
{1\over c} \left[ {1\over 3} N_a I^{(3)}_{ija} + {4\over 3}
\varepsilon_{ab(i} J^{(2)}_{j)a} N_b \right] \nonumber \\ &&\qquad
+{1\over c^2} \left[ {1\over 12} N_{ab} I^{(4)}_{ijab} + {1\over 2}
\varepsilon_{ab(i} J^{(3)}_{j)ac} N_{bc} \right] \nonumber \\ &&\qquad
+ {1\over c^3} \left[ {1\over 60} N_{abc} I^{(5)}_{ijabc} + {2\over
15} \varepsilon_{ab(i} J^{(4)}_{j)acd} N_{bcd} \right] \nonumber \\
&&\qquad +{1\over c^4} \left[ {1\over 360} N_{abcd} I^{(6)}_{ijabcd} +
{1\over 36} \varepsilon_{ab(i} J^{(5)}_{j)acde} N_{bcde} \right]
\nonumber \\ &&\qquad +{1\over c^5} \left[{1\over 2520} N_{abcde}
I^{(7)}_{ijabcde} + {1\over 210} \varepsilon_{ab(i} J^{(6)}_{j)acdef}
N_{bcdef} \right]\biggr\}\,,
\end{eqnarray}
where all the source moments are evaluated at the current time
$T_R$. The type (r) is
\begin{eqnarray}\label{eq:WFinst(r)}
\bigl(h_{km}^{\rm TT}\bigr)_{\rm inst(r)}&=& \frac{2G}{c^4 R}\,{\cal
P}_{ijkm} \frac{G}{c^5} \biggl\{ {1\over7}I^{(5)}_{a<i}I_{j>a}-{5
\over7}I^{(4)}_{a<i}I^{(1)}_{j>a}-{2
\over7}I^{(3)}_{a<i}I^{(2)}_{j>a}+{1
\over3}\varepsilon_{ab<i}I^{(4)}_{j>a}J_{b}\nonumber\\
&&\qquad+{1\over12}N_{ab}\left[-{21 \over5}I^{(5)}_{<ij}I_{ab>}-{63
\over5}I^{(4)}_{<ij}I^{(1)}_{ab>}-{102
\over5}I^{(3)}_{<ij}I^{(2)}_{ab>}\right]\nonumber\\
&&\qquad+{1\over2}N_{bc}\,\varepsilon_{abi}\left[
{1\over10}\varepsilon_{pq<j}I^{(5)}_{a\underline{p}}I_{c>q}
-{1\over2}\varepsilon_{pq<j}I^{(4)}_{a\underline{p}}
I^{(1)}_{c>q}-2J_{<j}I^{(4)}_{ac>}\right]\biggr\}\,.
\end{eqnarray}
Apart from two terms involving the source dipole moment $J_i$ or
angular momentum, these terms are made of quadrupole-quadrupole
couplings coming from $U_{ij}$, $U_{ijk}$ and $V_{ijk}$ in
Eqs.~(\ref{eq:U})--(\ref{eq:V}) and computed in Ref. \cite{Bquad}
(though, using dimensional and parity arguments, their structure can
easily be written down, the computation of the numerical coefficients
in front of each inst(r) term needs a detailed study). The inst(c)
terms refer to the instantaneous terms in the ``canonical'' moment
and can be written down as
\begin{eqnarray}\label{eq:WFinst(c)}
\bigl(h_{km}^{\rm TT}\bigr)_{\rm inst(c)}&=& \frac{2G}{c^4 R}\,{\cal
P}_{ijkm} \frac{G}{c^5} \biggl\{
4\left[W^{(4)}I_{ij}+W^{(3)}I_{ij}^{(1)}-W^{(2)}I_{ij}^{(2)}
-W^{(1)}I_{ij}^{(3)}\right]\biggr\}\,,
\end{eqnarray}
where $W$ is the particular ``monopole'' moment introduced in
(\ref{MijIij}). Both the inst(r) and inst(c) terms represent new
features of the 2.5PN waveform. Concerning hereditary parts, we have
the tail integrals (dominantly 1.5PN) which read as
\begin{eqnarray}\label{eq:WFtail}
\bigl(h_{km}^{\rm TT}\bigr)_{{\rm tail}}&=& \frac{2G}{c^4 R}{\cal
P}_{ijkm}{2G\,M \over c^3} \int_{-\infty}^{T_R} dV \left\{ \left[ \ln
\left({T_R-V\over 2b}\right) + {11\over 12}\right] I^{(4)}_{ij}(V)
\right.\nonumber\\ &&\qquad\left.+{N_a\over 3c}\left[ \ln \left(
{T_R-V\over 2b}\right)+{97\over60}\right] I^{(5)}_{ija} (V)
\right.\nonumber\\ &&\qquad\left. +{4 N_b\over 3c} \left[\ln \left(
{T_R-V\over 2b}\right)+{7\over6}\right]\varepsilon_{ab(i}
J^{(4)}_{j)a}(V)\right.\nonumber\\ &&\qquad\left.+{N_{ab}\over 12c^2}
\left[ \ln \left({T_R-V\over 2b}\right) + {59\over 30}\right]
I^{(6)}_{ijab}(V)\right.\nonumber\\ &&\qquad\left. +{N_{bc} \over
2c^2} \left[ \ln \left({T_R-V\over 2b}\right) + {5\over
3}\right]\varepsilon_{ab(i}J^{(5)}_{j)ac}(V) \right\}\,,
\end{eqnarray} 
and the non-linear memory integral (which is purely of order 2.5PN)
given by
\begin{equation}\label{eq:WFmemory}
\bigl(h_{km}^{\rm TT}\bigr)_{{\rm memory}} = \frac{2G}{c^4 R}{\cal
P}_{ijkm} {G \over c^5} \int_{-\infty}^{T_R} dV \left\{
-{2\over7}I^{(3)}_{a<i}(V)I^{(3)}_{j>a}(V)+{N_{ab} \over
30}I^{(3)}_{<ij}(V)I^{(3)}_{ab>}(V) \right\}\,.
\end{equation}
The latter expression is in complete agreement with the results of
\cite{WW-chris,BD92,Bquad}.

\subsection{Source multipole moments required at the 2.5PN order}
\label{sec:multipole moments}

Evidently the above formulas remain empty unless we feed them with the
explicit expressions of the source multipole moments, essentially the
mass-type $I_L$ and current-type $J_L$, appropriate for a specific
choice of matter model. In the present Section, we list the $I_L$'s
and $J_L$'s needed for the 2.5PN accurate waveform in the case of
point particles binaries in circular orbits. This is the extension of
the list of moments given in Eqs. (4.4) of \cite{BDI} for the
computation of the 2PN accurate waveform. We do not give any details
on this calculation because it follows exactly the same techniques as
in Ref. \cite{BIJ}.

Up to 2.5PN order in the waveform the mass moments are
{\allowdisplaybreaks
\begin{subequations}
\begin{eqnarray}\label{eq:massmom}
I_{ij} &=& \nu\,m\, {\rm STF}_{ij} \left\{{x}^{ij}\left[1 + \gamma
\left(-{1\over 42}-{13\over 14}\nu \right) + \gamma^2 \left(-{461\over
1512} -{18395\over 1512}\nu - {241\over 1512} \nu^2\right)\right]
\right.\nonumber\\ &&~~\left. +{r^2\over c^2}{v}^{ij}\left[{11\over
21}-{11\over 7}\nu + \gamma \left({1607\over 378}-{1681\over 378} \nu
+{229\over 378}\nu^2\right)\right] + {48\over 7}\,{r\over c}
x^{i}v^{j}\nu\gamma^2\right\}\nonumber\\ &&~~+{\cal O}(6) \,,\\
I_{ijk} &=& \nu\,m\,( X_2 - X_1 )\, {\rm STF}_{ijk} \left\{ {x}^{ijk}
\left[1 -\gamma \nu- \gamma^2 \left({139\over 330}+{11923\over 660}\nu
+{29\over 110}\nu^2\right) \right]\right.\nonumber\\ &&~~+{r^2\over
c^2}x^{i}v^{jk} \left[1 - 2\nu - \gamma \left(-{1066\over
165}+\left.{1433\over 330}\nu -{21\over 55} \nu^2\right)
\right]\right\}\,\nonumber\\ &&~~+{\cal O}(5)\,,\\ I_{ijkl} &=& \nu
\,m\, {\rm STF}_{ijkl}\, \left\{ x^{ijkl}\left[1 - 3\nu + \gamma
\left({3\over 110} - {25\over 22}\nu + {69\over
22}\nu^2\right)\right]\right.\nonumber\\
&&~~+\left.\frac{78}{55}\,\frac{r^2}{c^2}\,v^{ij} x^{kl} ( 1 - 5\nu +
5\nu^2 ) \right\} \,+{\cal O}(4)\,, \\ I_{ijklm}&=&\nu\,m\,(X_2-X_1)\,
{\rm STF}_{ijklm} \left\{ x^{ijklm}\left[1-2\nu+\gamma\left({2\over
39}-{47\over 39}\nu+{28\over 13}\nu^2 \right)\right]\right.\nonumber
\\ &&~~+\left. \frac{70}{39}\,{r^2 \over c^2}
x^{ijk}v^{lm}\left(1-4\nu+3\nu^2\right)\right\} \,+{\cal O}(3)\,,\\
I_{ijklmn}&=&\nu\,m\,{\rm STF}_{ijklmn} \left\{ x^{ijklmn}(1-5 \nu+5
\nu^2)\right\} \, +{\cal O}(2)\,,\\ I_{ijklmno}&=& \nu\,m\,(X_2-X_1)\
(1-4 \nu + 3 \nu^2)\,{\rm STF}_{ijklmno}
\left\{x^{ijklmno}\right\}+{\cal O}(1)\,.
\end{eqnarray}\end{subequations}}\noindent
Further, the current moments are given by
{\allowdisplaybreaks
\begin{subequations}\label{eq:currentmom}
\begin{eqnarray}
J_{ij} &=& \nu\,m\, (X_2-X_1)\,{\rm STF}_{ij}\left\{ \varepsilon_{abi}
x^{ja}v^b \left[1 +\gamma \left({67\over 28}-{2\over 7}\nu
\right)\right.\right.\nonumber \\
&&~~\left.\left. +\gamma^2\left({13\over 9} -{4651\over 252}\nu
-{1\over 168}\nu^2 \right)\right]\right\} \,+{\cal O}(5)\,, \\
J_{ijk} &=& \nu\,m \,{\rm STF}_{ijk}\left\{ \varepsilon_{kab} x^{aij}
v^b \left[1 - 3\nu + \gamma \left({181\over 90} - {109\over 18}\nu +
{13\over 18}\nu^2\right)\right] \right.\nonumber \\
&&~~+\left.\frac{7}{45}\,\frac{r^2}{c^2}\varepsilon_{kab}x^a v^{bij}(1
- 5\nu + 5\nu^2)\right\}\,+{\cal O}(4)\,,\\ J_{ijkl} &=& \nu\,m \,
(X_2 - X_1) \,{\rm STF}_{ijkl} \left\{ \varepsilon_{lab}x^{ijka} v^b
\left[1-2\nu \right.\right. \nonumber \\ &&\left.\left. +\gamma
\left({20\over 11}-{155\over 44}\nu+{5\over 11}\nu^2\right) \right]
+\frac{4}{11}\,{r^2\over c^2}\varepsilon_{lab}x^{ia}
v^{jkb}\left(1-4\nu+3\nu^2\right)\right\}\,\nonumber \\ &&~~ +{\cal
O}(3)\,,
\label{eq:4.4h}\\ J_{ijklm} &=& \nu\,m \, {\rm STF}_{ijklm} \left\{
\varepsilon_{mab}x^{aijkl}v^b \left(1-5\nu + 5\nu^2\right)
\right\}+{\cal O}(2)\,,\\ J_{ijklmn} &=& \nu\,m \, (X_2-X_1) (1-4\nu +
3\nu^2)\,{\rm STF}_{ijklmn} \left\{ \varepsilon_{nab}
x^{aijklm}v^{b}\right\}+{\cal O}(1)\,.
\end{eqnarray}
\end{subequations}}\noindent
[We recall that $X_1=\frac{m_1}{m}$, $X_2=\frac{m_2}{m}$, and
$\nu=X_1X_2$; the PN parameter $\gamma$ is defined by (\ref{gam}); the
STF projection is mentioned explicitly in front of each term.]

In addition, the current dipole $J_i$ in (\ref{eq:WFinst(r)}) is the
constant binary's total angular momentum which needs to be given only
at Newtonian order, and we need also to give the monopolar moment $W$
which appears inside the inst(c) terms of (\ref{eq:WFinst(c)}) and
comes from the relation (\ref{MijIij}) between canonical and source
quadrupoles. We have
\begin{subequations}\label{JW}\begin{eqnarray}
J_i &=& \nu\,m \, \varepsilon_{iab}x^av^b + {\cal O}(2)\,
,\label{dipoleJ}\\ W &=& {1\over 3} \, \nu\,m \, {\bf x}.{\bf v} +
{\cal O}(2)\,.\label{monopoleW}
\end{eqnarray}\end{subequations}
With all the latter source moments valid for a specific matter system
(compact binary in circular orbit) the gravitational waveform is fully
specified up to the 2.5PN order.

\section{Hereditary terms at the 2.5PN order}\label{sec:hered}

We now come to the computation of the hereditary terms, {\it i.e.}
made of integrals extending on all the past history of the
non-stationary source, at any time $V$ from $-\infty$ in the past up
to $T_R=T-R/c$. In the following we shall refer to $T_R$ as the {\it
current} time --- the one at which the observation of the radiation
field occurs. As we have seen in Section \ref{insthered}, at the 2.5PN
order the waveform contains two types of hereditary terms: tail
integrals, given by Eq.~(\ref{eq:WFtail}), and the non-linear memory
integral (\ref{eq:WFmemory}).\footnote{Notice that in the
gravitational-wave {\it flux} (in contrast to the waveform), the
non-linear memory integral is instantaneous --- it is made of a simple
time anti-derivative and the flux depends on the time-derivative of
the waveform.}

Evidently, in order to evaluate the hereditary terms, one must take
into account the fact that the binary's orbit will have evolved, by
gravitational radiation reaction, from early time on. However we shall
show, following Refs. \cite{BS-93,Wiseman}, that the tails can
basically be computed using the binary's {\it current} dynamics at
time $T_R$, {\it i.e.} a circular orbit travelled at the current
orbital frequency $\omega (T_R)$ (this will be true modulo negligible
4PN terms in the waveform). Concerning the memory integral one
definitely needs taking into account a model of binary evolution in
the past.

\subsection{Model for adiabatic inspiral in the remote past}
\label{model}

In this paper we adopt a simplified model of binary's past evolution in
which the orbit decays adiabatically because of gravitational
radiation damping according to the standard quadrupolar ({\it i.e.}
Newtonian) approximation. We shall justify later that such a model is
sufficient for our purpose --- because we shall have to take into
account the PN corrections in the tails only at the current epoch. The
orbit will be assumed to remain circular, apart from the gradual
inspiral, at any time $V<T_R$. We shall ignore any astrophysical
(non-gravitational) processes such as the binary's formation by
capture process in some dense stellar cluster, the successive supernova
explosions and associated core collapses leading to the formation of
the two compact objects, etc.

Let us recall the expressions of the binary's orbital parameters as
explicit functions of time $V$ in the quadrupolar circular-orbit
approximation \cite{peters}. The orbital separation $r(V)$ evolves
according to a power law, namely
\begin{equation}\label{rV}
r(V) = 4\left[\frac{G^3m^3\nu}{5\,c^5}(T_c-V)\right]^{1/4}\,,
\end{equation}
where $T_c$ denotes the coalescence instant, at which the two bodies
merge together and the orbital frequency formally tends to
infinity. The factor $1/c^5$ therein represents the 2.5PN order of
radiation reaction. We assume that our current detection of the binary
takes place before the coalescence instant, $T_R < T_c$, in a regime
where the binary inspiral is {\it adiabatic} and the approximation
valid.\footnote{The formal PN order of the time interval left till
coalescence is the inverse of the order of radiation-reaction, $T_c -
T_R={\cal O}(c^5)={\cal O}(-5)$.}

The orbital frequency $\omega$ in this model ($\omega=2\pi/P$, where
$P$ is the orbital period), is related at any time to the orbital
separation (\ref{rV}) by Kepler's (Newtonian) law
$G\,m=r^3\,\omega^2$. [Again we shall justify later our use of a
Newtonian model for the early-time inspiral.] Hence,
\begin{equation}\label{omV}
\omega(V) = \frac{1}{8}\left[\frac{G^{5/3}m^{5/3}\nu}{5\,c^5}(T_c-V)
\right]^{-3/8}\,.
\end{equation}
Instead of $\omega$ it is convenient to make use of the traditional
frequency-related post-Newtonian parameter $x\equiv (G\,
m\,\omega/c^3)^{2/3}$ already considered in (\ref{x}). It is given by
\begin{equation}\label{xV}
x(V) = \frac{1}{4}\left[\frac{\nu\,c^3}{5\,G\,m}(T_c-V)\right]^{-1/4}\,.
\end{equation}
The orbital phase $\phi=\int \omega\,dt=\frac{c^3}{G\,m}\int x^{3/2}
dt$, namely the angle between the binary's separation and the
ascending node ${\cal N}$, reads as
\begin{equation}
\phi(V)=\phi_c-\frac{1}{\nu}\left[\frac{\nu\,c^3}{5\,G\,m}(T_c-V)\right]^{5/8}\,,
\label{phiV}\end{equation}
where $\phi_c$ is the value of the phase at the coalescence instant.

The latter expressions are inserted into the various hereditary terms,
and integrated from $-\infty$ in the past up to now. In order to
better understand the structure of the integrals, it is advisable to
re-express the above quantities (\ref{rV})--(\ref{phiV}) in terms of
their values at the {\it current} time $T_R$. A simple way to achieve
this is to introduce, following \cite{BS-93}, the new time-related
variable
\begin{equation}
y\equiv\frac{T_R-V}{T_c-T_R}\,,
\label{yvar}\end{equation}
and to make use of the power-law dependence in time of
Eqs.~(\ref{rV})--(\ref{phiV}). This leads immediately, for the orbital
radius $r(V)$ and similarly for $\omega(V)$ and $x(V)$, to
\begin{equation}
r(V) = r(T_R)(1+y)^{1/4}\,,\label{rVy}
\end{equation}
where $r(T_R)$ refers to the current value of the radius. For the
orbital phase we get
\begin{equation}
\phi(V) = \phi(T_R)-\frac{1}{\nu}\left[\frac{\nu\,c^3}{5\,G\,m}(
T_c-T_R)\right]^{5/8}\Bigl[(1+y)^{5/8}-1\Bigr]\,.
\label{phiVy0}\end{equation}
The latter form is however not the one we are looking for. Instead we
want to make explicit the fact that the phase difference between $T_R$
and some early time $V$ will become larger when the inspiral rate gets
slower, {\it i.e.}  when the {\it relative} change of the orbital
frequency $\omega$ in one corresponding period $P$ becomes smaller.

To this end we introduce a dimensionless ``adiabatic parameter''
associated with the inspiral rate {\it at the current time}
$T_R$. This is properly defined as the ratio between the current
period and the time left till coalescence. We adopt the
definition\,\footnote{We have $\xi=\frac{8}{3}\,{\dot
\omega}/\omega^2$ so our definition agrees with the actual relative
frequency change $\propto {\dot \omega}/\omega^2$ in one period. It is
also equivalent to the one adopted in \cite{BS-93}: $\xi=\xi_{\rm
BS}/\pi$.}
\begin{equation}
\xi(T_R)\equiv \frac{1}{(T_c-T_R)\,\omega(T_R)}\,.
\label{xiTR}\end{equation}
The adiabatic parameter $\xi$ is of the order 2.5PN. Written in terms
of the PN variable $x$ defined by (\ref{xV}) it reads
\begin{equation}
\xi(T_R) = \frac{256\,\nu}{5}\,x^{5/2}(T_R)\,.
\label{xiTR'}\end{equation}
Now Eq. (\ref{phiVy0}) can be expressed with the help of $\xi(T_R)$ in
the more interesting form
\begin{equation}
\phi(V) = \phi(T_R)-\frac{8}{5\,\xi(T_R)}\Bigl[(1+y)^{5/8}-1\Bigr]\,,
\label{phiVy}\end{equation}
which makes it clear that the phase difference $\Delta\phi = \phi(T_R)
- \phi(V)$, which is $2\pi$ times the number of orbital cycles between
$V$ and $T_R$, tends to infinity when $\xi(T_R)\rightarrow 0$ on any
``remote-past'' time interval for which $y$ is bounded from below, for
instance $y>1$. What is important is that (\ref{phiVy}) depends on the
{\it current} value of the adiabatic parameter, so we shall be able to
compute the hereditary integrals in the relevant limit where
$\xi(T_R)\rightarrow 0$, appropriate to the current adiabatic
regime. Notice that, at recent time, when $V\rightarrow T_R$ or
equivalently $y\rightarrow 0$, we have
\begin{equation}
\phi(V) = \phi(T_R)-\frac{y}{\xi(T_R)}+\mathcal{O}\left(
y^2\right)\,,
\label{phiVlimit}\end{equation}
which is of course the same as the Taylor expansion 
\begin{equation}\phi(V) =
\phi(T_R)-(T_R-V)\,\omega(T_R)+\mathcal{O}\left[(T_R-V)^2\right]\,.
\label{phiVlimiteq}\end{equation}

\subsection{The nonlinear memory integral}\label{sec:NLM}

We tackle the computation of the novel hereditary term at the 2.5PN
order, namely the non-linear memory integral given by
(\ref{eq:WFmemory}). As we shall see the computation boils down to the
evaluation of only two ``elementary'' hereditary integrals, below
denoted $I(T_R)$ and $J(T_R)$. A third type of elementary integral,
$K(T_R)$, will be necessary to compute the tail integrals in Section
\ref{sec:tail}.

The two wave polarisations corresponding to Eq.~(\ref{eq:WFmemory}),
calculated with our conventions and notation explained after
(\ref{plus-cross}), are readily obtained from the Newtonian
approximation to the quadrupole moment $I_{ij}$ [first term in
(\ref{eq:massmom})], and cast into the form
\begin{subequations}\label{plus-crossmem}
\begin{eqnarray}
\bigl(h_+ \bigr)_{{\rm memory}} (T_R) &=& \frac{2G}{c^4
R}\frac{G^4m^5\nu^2\sin^2i}{c^5}\int_{-\infty}^{T_R} \frac{dV}{r^5(V)}
\left\{-\frac{12}{5}+\frac{2}{15}\sin^2i\right.
\label{plusmem}\nonumber\\&&\qquad\qquad\qquad\qquad\quad\left.
+\left(\frac{4}{15}-\frac{2}{15}\sin^2i\right)\cos
[4\phi(V)]\right\}\,,\\ \bigl(h_\times \bigr)_{{\rm memory}} (T_R)
&=& \frac{2G}{c^4 R}\frac{G^4m^5\nu^2\sin^2i}{c^5}\int_{-\infty}^{T_R}
\frac{dV}{r^5(V)}\left\{\frac{4 \cos i}{15} \sin [4\phi(V)]\right\}\,.
\label{crossmem}\end{eqnarray}
\end{subequations}
In our model of binary evolution the radius of the orbit, $r(V)$, and
orbital phase, $\phi (V)$, are given by Eqs.~(\ref{rV}) and
(\ref{phiV}), at any time such that $V < T_R < T_c$. Because
$r(V)\propto (T_c-V)^{1/4}$ we see that the integrals in
(\ref{plus-crossmem}) are perfectly well-defined, and in fact
absolutely convergent at the bound $V\rightarrow -\infty$. There are
two distinct types of terms in (\ref{plus-crossmem}). A term, present
only in the {\it plus} polarisation (\ref{plusmem}), is independent of
the orbital phase $\phi$, and given by a steadily varying function of
time, having an amplitude increasing like some power law but without
any oscillating behaviour. This ``steadily increasing'' term is
specifically responsible for the memory effect. The other terms,
present in both polarisations, oscillate with time like some sine or
cosine of the phase, in addition of having a steadily increasing
maximal amplitude.

Consider first the steadily growing, non-oscillating term. Its
computation simply relies, as clear from (\ref{plusmem}), on the
single elementary integral
\begin{equation}
I(T_R) \equiv \frac{(G\,m)^4}{c^7} \int_{-\infty}^{T_R}
\frac{dV}{r^5(V)}\,,
\label{intI}\end{equation}
where we find convenient to factorize out an appropriate coefficient
in order to make it dimensionless. The calculation of (\ref{intI}) is
easily done directly, but it is useful to perform our change of
variable (\ref{yvar}), as an exercise to prepare the treatment of the
(somewhat less easy) oscillating terms. Thus, we write in a first
stage
\begin{equation}
I(T_R) = \frac{(G\,m)^4}{c^7}\,\frac{T_c-T_R}{r^5(T_R)}
\int_0^{+\infty} \frac{dy}{(1+y)^{5/4}}\,.
\label{intIy}\end{equation}
The factor in front is best expressed in terms of the dimensionless PN
parameter $x(T_R)$, and of course the remaining integral is trivially
integrated. We get
\begin{equation}
I(T_R) = \frac{5}{256\,\nu}\,x(T_R) \int_0^{+\infty}
\frac{dy}{(1+y)^{5/4}} = \frac{5}{64\,\nu}\,x(T_R)\,.
\label{intIres}\end{equation}

With this result we obtain the steadily-increasing or memory term in
Eq.~(\ref{plusmem}). However, as its name indicates, this term keeps a
``memory" of the past activity of the system. As a test of the
numerical influence of the binary's past history on
Eq.~(\ref{intIres}), let us suppose that the binary was created {\it
ex nihilo} at some finite initial time $T_0$ in a circular orbit
state. This premise is not very realistic --- we should more
realistically assume {\it e.g.} a binary capture in some stellar
cluster and/or consider an initially eccentric orbit --- but it should
give an estimate of how sensitive is the memory term on initial
conditions. In this crude model we have to consider the integral
$I_0(T_R)$ extending from $T_0$ up to $T_R$. We find that the ratio of
$I_0(T_R)$ and our earlier model $I(T_R)$ is
\begin{equation}
\frac{I_0(T_R)}{I(T_R)} = 1 -
\left(\frac{T_c-T_R}{T_c-T_0}\right)^{1/4} = 1 -
\frac{r(T_R)}{r(T_0)}\,.
\label{IinI}\end{equation}
Let us choose the observation time $T_R$ such that the binary is
visible by the VIRGO/LIGO detectors. At the entry of the detectors'
frequency bandwidth, say $\omega_{\rm seismic}\simeq 30\,$Hz, we
obtain $T_c-T_R\simeq 10^3\,$s and $r(T_R)\simeq 700\,$km in the case
of two neutron stars ($m=2.8\,M_{\odot}$). For a binary system
initially formed on an orbit of the size of the Sun, $r(T_0)\simeq
10^6\,$km (corresponding to $T_c-T_0\simeq 10^8\,$yr), we find that
the fractional difference between our two models $I_0$ and $I$ amounts
to about $10^{-3}$. For an initial orbit of the size of a white dwarf,
$r(T_0)\simeq 10^4\,$km, the fractional difference is of the order of
10\% --- rather large indeed. So we conclude that indeed the memory
term in (\ref{plusmem}), depends rather severely on detailed
assumptions concerning the past evolution of the binary system. We
shall have to keep this feature in mind when we present our final
results for this term.

Turn next our attention to the phase-dependent, oscillating terms in
Eqs.~(\ref{plus-crossmem}). Clearly these terms are obtained once we
know the elementary integral
\begin{equation}
J(T_R) \equiv \frac{(G\,m)^4}{c^7} \int_{-\infty}^{T_R} dV\,\frac{e^{4i
\phi(V)}}{r^5(V)}\,.
\label{intJ}\end{equation}
Inserting (\ref{rVy}) and (\ref{phiVy}) into it we are led to the form
[which exactly parallels (\ref{intIres})]
\begin{equation}
J(T_R) = \frac{5}{256\,\nu}\,x(T_R) \, e^{4i \phi(T_R)}
\int_0^{+\infty} \frac{dy}{(1+y)^{5/4}}\,
e^{-\frac{32i}{5\xi(T_R)}\left[(1+y)^{5/8}-1\right]}\,.
\label{intJy}\end{equation}
We shall compute this integral in the form of an approximation series,
valid in the adiabatic limit $\xi (T_R)\rightarrow 0$. The easiest way
to obtain successive approximations is to integrate by parts. We
obtain
\begin{equation}
\int_0^{+\infty} \frac{dy}{(1+y)^{5/4}}
e^{-\frac{32i}{5\xi}\left[(1+y)^{5/8}-1\right]}=\frac{\xi}{4i}\left\{
1-\frac{7}{8}\int_0^{+\infty} \frac{dy}{(1+y)^{15/8}}
e^{-\frac{32i}{5\xi}\left[(1+y)^{5/8}-1\right]}\right\}\,.
\label{intparts}\end{equation}
A further integration by parts shows that the integral in the curly
brackets of (\ref{intparts}) is itself of order $\xi$, so we have the
result\,\footnote{By successive integration by parts one generates the
asymptotic series (divergent for any value of $\xi$)
$$\int_0^{+\infty} \frac{dy}{(1+y)^{5/4}}
e^{-\frac{32i}{5\xi}\left[(1+y)^{5/8}-1\right]}\,\sim\,-8\sum_{n=1}^{
+\infty}(5n-3)(5n-8)\cdots (12)(7)\left(\frac{i\xi}{32}\right)^n\,,$$
where we have used the standard notation $\sim$ for equalities valid
in the sense of asymptotic series.}
\begin{equation}
\int_0^{+\infty} \frac{dy}{(1+y)^{5/4}}
e^{-\frac{32i}{5\xi}\left[(1+y)^{5/8}-1\right]}
=\frac{\xi}{4i}\Bigl\{1+\mathcal{O} \left(\xi\right)\Bigr\}\,.
\label{intpartsapprox}\end{equation}
A standard way to understand it is to remark that, when $y$ is
different from zero, the phase of the integrand of
(\ref{intpartsapprox}) oscillates very rapidly when $\xi\rightarrow
0$, so the integral is made of a sum of alternatively positive and
negative terms and is essentially zero. Consequently, the value of the
integral is essentially given by the contribution due to the bound at
$y=0$, which can be approximated by
\begin{equation}\label{argument}
\int_{0} dy\,e^{-\frac{4iy}{\xi}} = \frac{\xi}{4i}\,.
\end{equation}

Because $\xi(T_R)$ is of order 2.5PN the result (\ref{intpartsapprox})
is sufficient for the control of the 2.5PN waveform, thus our
elementary integral reads, with the required precision,

\begin{equation}
J(T_R) = x^{7/2}(T_R)\,\frac{e^{4i\phi
(T_R)}}{4i}\Bigl\{1+\mathcal{O}\left(\xi\right)\Bigr\}\,.
\label{intJres}\end{equation}
We find that the ``oscillating'' integral $J(T_R)$ is an order 2.5PN
smaller than the ``steadily growing'' or ``memory'' integral
$I(T_R)$. This can be interpreted by saying that the cumulative
(secular) effect of the integration over the whole binary inspiral in
$I(T_R)$ is comparable to the inverse of the order of radiation
reaction forces --- a quite natural result. By contrast the
oscillations in $J$, due to the sequence of orbital cycles in the
entire life of the binary system, compensate more or less each other
yielding a net result which is 2.5PN smaller than for
$I$. Furthermore, the argument leading to the evaluation of
(\ref{argument}) shows that $J$, contrarily to $I$, is quite
insensitive to the details of the binary's past evolution.

Substituting Eqs.~(\ref{intIres}) and (\ref{intJres}) into
(\ref{plus-crossmem}) we finally obtain the hereditary memory-type
contributions to the polarisation waveforms as
\begin{subequations}\label{plus-crossresmem}
\begin{eqnarray}\label{plus-resmem}
\bigl(h_+ \bigr)_{{\rm memory}} &=& \frac{2\,G \,m \,\nu \,x}{c^2 R}
\sin^2i \left\{ -\frac{17+\cos^2i}{96}+\frac{\nu}{30} x^{5/2}
(1+\cos^2i) \sin (4\phi) \right\}\,,\nonumber\\\\ \bigl(h_\times
\bigr)_{{\rm memory}} &=& \frac{2\,G \,m \,\nu \,x}{c^2 R} \sin^2i
\left\{ -\frac{\nu}{15}x^{5/2} \cos i \,\cos (4\phi) \right\}\,,
\end{eqnarray}
\end{subequations}
where of course all quantities correspond to the current time $T_R$.
The phase-independent term is the non-linear memory or Christodoulou
effect \cite{chris,WW-chris,Thorne-chris,BD92,Bquad} in the case of
inspiralling compact binaries. Our calculation, leading to a factor
$\propto\sin^2i(17+\cos^2i)$, agrees with the result of Wiseman and
Will \cite{WW-chris}.\footnote{The difference of a  factor of $-2$ of between
their and our result is here probably due to a different convention for the
polarisation waveforms.} The non-linear memory term {\it
stricto-sensu} affects the plus polarisation but not the cross
polarisation, for which it is exactly {\it zero}. As we have said, it
represents a part of the waveform whose amplitude grows with time, but
which is nearly constant over one orbital period. It is therefore
essentially a {\it zero-frequency} (DC) effect, which has rather poor
observational consequences in the case of the LIGO-VIRGO detectors,
whose frequency bandwidth is always limited from below by $\omega_{\rm
seismic}>0$. In addition, we know that detecting and analyzing the
ICBs relies essentially on monitoring the phase evolution, which in
turn is determined by the total gravitational-wave flux. But we have
already noticed that the non-linear memory integral
(\ref{eq:WFmemory}) is instantaneous in the flux (and in fact it does
not contribute to the phase in the circular orbit case
\cite{Beflux-96}). It seems thus that the net cumulative ``memory''
change in the waveform of ICBs is hardly detectable. On the other
hand, the frequency-dependent terms found in
Eq.~(\ref{plus-crossresmem}) form an integral part of the 2.5PN
waveform.

An important comment is in order. As we have seen the memory effect,
{\it i.e.} the DC term in Eq. (\ref{plus-resmem}), is ``Newtonian''
because the cumulative integration over the binary's past just
compensates the formal 2.5PN order of the hereditary integral. Thus we
expect that some formal PN terms strictly higher that 2.5PN will
actually contribute to the 2.5PN waveform {\it via} a similar
cumulative integration. For instance at the 3.5PN level there will be
a memory-type integral in the radiative quadrupole moment $U_{ij}$,
which is quadratic in the mass octupole moment (of the symbolic form
$I_{ab<i}\times I_{j>ab}$). After integration over the past using our
model of adiabatic inspiral, we expect that the ``steadily-growing''
part of the integral should yield a DC contribution to the waveform at
the relative 1PN order. In the present paper we do not consider such
higher-order post-Newtonian DC contributions, and leave their
computation to future work.

\subsection{Gravitational-wave tails}\label{sec:tail}

The tails up to 2.5PN order are given by
Eq.~(\ref{eq:WFtail}). Because of the logarithmic kernel they involve,
the tails are more complicated than simple time anti-derivatives, and
they constitute a crucial part of both the waveform and the energy
flux, and in particular of the orbital phase with important
observational consequences (for a review see \cite{Blanchet-rev}).

The computation of tails reduces to the computation of a new type of
``elementary'' integral, differing from $J(T_R)$ given in (\ref{intJ})
by the presence of an extra logarithmic factor in the integrand, and
given by\,\footnote{Actually in order to describe the tails we should
consider the more general integrals
$$K_{n,p}(T_R) = \frac{(G\,m)^{p-1}}{c^{2p-3}} \int_{-\infty}^{T_R}
dV\,\frac{e^{i n
\phi(V)}}{r^p(V)}\,\ln\left(\frac{T_R-V}{T_c-T_R}\right)\,.$$ The
dominant tails at the 1.5PN order correspond to $p=4$ and $n=2$, the
tails at the 2PN order have $p=9/2$ and $n=1,3$, and those at the
2.5PN order have $p=5$ and $n=2,4$. Here we deal with the particular
case $p=5$, $n=4$ because the calculation parallels the one of
$I(T_R)$ and $J(T_R)$ in Section \ref{sec:NLM}. The other cases are
treated in a similar way.}
\begin{equation}
K(T_R) \equiv \frac{(G\,m)^4}{c^7} \int_{-\infty}^{T_R}
dV\,\frac{e^{4i
\phi(V)}}{r^5(V)}\,\ln\left(\frac{T_R-V}{T_c-T_R}\right)\,.
\label{intK}\end{equation}
For convenience we have inserted into the logarithm of (\ref{intK})
the constant time scale $T_c-T_R$ instead of the normalization $2b$
more appropriate for tails [or, for instance, $2b\,e^{-11/12}$, see
(\ref{eq:WFtail})], but we can do this at the price of adding another
term which will be proportional to $J(T_R)$ already computed in
Section \ref{sec:NLM}.

With the help of the $y$-variable (\ref{yvar}) we transform the latter
integral into
\begin{equation}
K(T_R) = \frac{5}{256\,\nu}\,x(T_R) \, e^{4i \phi(T_R)}
\int_0^{+\infty} \frac{dy\,\ln y}{(1+y)^{5/4}}\,
e^{-\frac{32i}{5\xi(T_R)}\left[(1+y)^{5/8}-1\right]}\,.
\label{intKy}\end{equation}
Because of the factor $\ln y$ we are not able to directly integrate by
parts as we did to compute $J(T_R)$. Instead we must split the
integral into some ``recent-past'' contribution, say $y\in
\bigr]0,1\bigl[$, and the ``remote-past'' one, $y\in
\bigr]1,+\infty\bigl[$. In the remote-past integral, whose lower bound
at $y=1$ allows for differentiating the factor $\ln y$, we perform the
integrations by parts in a way similar to (\ref{intparts}). This leads
to
\begin{equation}
\int_1^{+\infty} \frac{dy\,\ln y}{(1+y)^{5/4}}\,
e^{-\frac{32i}{5\xi}\left[(1+y)^{5/8}-1\right]} = \mathcal{O}
\left(\xi^2\right)\,,
\label{remote}\end{equation}
which is found to be of the order of $\mathcal{O} \left(\xi^2\right)$
instead of $\mathcal{O} \left(\xi\right)$ in
Eq.~(\ref{intpartsapprox}). This is a consequence of the $\ln y$ which
is zero at the bound $y=1$, and thus kills the all-integrated
term. Hence we deduce from (\ref{remote}) that the contribution from
the remote past in the tail integrals is in fact quite small. The
details concerning the remote-past activity of the source are
negligible when computing the tails. More precisely, because
$\xi=\mathcal{O}(5)$, we can check that terms such as (\ref{remote})
do not contribute to the waveform before the 4PN order, so the tail
integrals can be legitimately approximated, with 2.5PN accuracy, by
their ``recent-past'' history (in agreement with the findings of
\cite{BS-93,Wiseman}).

Now, in the recent-past integral, {\it i.e.} $y\in \bigr]0,1\bigl[$,
we are allowed to replace the integrand by its equivalent when
$y\rightarrow 0$, modulo terms of the same magnitude as
(\ref{remote}). This fact has been proved rigorously in the Appendix B
of \cite{BS-93}. Here we shall not reproduce the proof but simply
state the end result, which reads
\begin{equation}
\int_0^1 \frac{dy\,\ln y}{(1+y)^{5/4}}\,
e^{-\frac{32i}{5\xi}\left[(1+y)^{5/8}-1\right]} = \int_0^1 dy\,\ln y
\, e^{-\frac{4i\,y}{\xi}} + \mathcal{O} \left(\xi^2\ln\xi\right)\,.
\label{recent}\end{equation}
As we see, the remainder is $\mathcal{O} \left(\xi^2\ln\xi\right)$,
instead of being merely $\mathcal{O} \left(\xi^2\right)$ as in
Eq.~(\ref{remote}). The integral in the R.H.S. of (\ref{recent}) gives
the main contribution to the tail integral (the only one to be taken
into account up to very high 4PN order). It is easily computed by
using a standard formula,\footnote{For any real number $\epsilon$ we
have
$$\epsilon\int_0^1 dy\,\ln y \, e^{i\epsilon y} + i\,\int_1^{+\infty}
\frac{dy}{y}\, e^{i\epsilon y} = -\frac{\pi}{2}\,{\rm
sgn}(\epsilon)-i\,\bigl(\ln |\epsilon | + C \bigr)\,,$$ where ${\rm
sgn}(\epsilon)$ and $|\epsilon |$ denote the sign and absolute value
of $\epsilon$.}
\begin{equation}
\int_0^1 dy\,\ln y \, e^{-\frac{4i\,y}{\xi}} =
\frac{\xi}{4i}\left[\frac{\pi}{2i} - \ln \left(\frac{4}{\xi}\right) -
C \right]+ \mathcal{O} \left(\xi^2\right) \,,
\label{recent'}\end{equation}
where $C=0.577\cdots$ is the Euler constant.

Finally our needed result is
\begin{equation}
K(T_R) = x^{7/2}(T_R)\,\frac{e^{4i\phi
(T_R)}}{4i}\left\{\frac{\pi}{2i} - \ln \left(\frac{4}{\xi(T_R)}\right)
- C +\mathcal{O}\left(\xi\ln\xi\right)\right\}\,,
\label{intKres}\end{equation}
where it is crucial that all the binary's parameters are evaluated at
the current time $T_R$. It is interesting to compare (\ref{intKres})
with our earlier result for $J(T_R)$ given by (\ref{intJres}), in
order to see the effect of adding a logarithmic-type kernel into an
oscillating, phase-dependent integral. Eq.~(\ref{intKres}), together
with its trivial extension to $K_{n,p}$, is used for the computation
of all the tails in the waveform at the 2.5PN order (and it could in
fact be used up to the order 3.5PN included). Actually we still have
to justify this because during the derivation of (\ref{intKres}) we
employed a {\it Newtonian} model for the binary's inspiral in the
past, {\it e.g.} we assumed the Kepler law $G\,m=r^3(V)\,\omega^2(V)$
at any time $V < T_R$. In the case of the memory term
(\ref{eq:WFmemory}) this is okay because it needs to be evaluated with
Newtonian accuracy. But in the case of tails, Eq.~(\ref{eq:WFtail}),
the dominant effect it at the 1.5PN order, so we have to take into
account a 1PN relative correction in order to control the 2.5PN
waveform. Nevertheless, our model of Newtonian inspiral in the past
{\it is} compatible with taking into account 1PN effects, for the
basic reason that for tails the past behaviour of the source is
negligible, so the 1PN effects have only to be included into the {\it
current} values of the parameters, {\it i.e.} $x(T_R)$, $\phi (T_R)$
and $\xi (T_R)$, in Eq.~(\ref{intKres}).

To see more precisely how this works, suppose that we want to replace
the ``Newtonian'' inspiral (\ref{rV}) by the more accurate 1PN law,
\begin{equation}\label{rV1PN}
r(V) =
4\left[\frac{G^3m^3\nu}{5c^5}(T_c-V)\right]^{1/4}+\frac{G\,m\,\eta}{c^2}
\,,
\end{equation}
where $\eta\equiv-\frac{1751}{1008}-\frac{7}{12}\nu$ (see {\it e.g.}
\cite{Blanchet-rev}). Substituting the variable $y$ we obtain the 1PN
equivalent of (\ref{rVy}) as
\begin{equation}
r(V) = r(T_R)(1+y)^{1/4}\biggl\{1+\eta\Bigl[(1+y)^{-1/4}-1\Bigr]
x(T_R)\biggr\}\,,
\label{rVy1PN}
\end{equation}
where we parametrized the 1PN correction term by means of the variable
$x$ evaluated at {\it current} time $T_R$ (we consistently neglect
higher PN terms). For the orbital phase we get

\begin{equation}
\phi(V) = \phi(T_R)-\frac{8}{5\,\xi(T_R)}\biggl\{\Bigl[(1+y)^{5/8}
-1\Bigr]\Bigl(1+\zeta\,x(T_R)\Bigr)
+\tau\Bigl[(1+y)^{3/8}-1\Bigr]x(T_R)\biggr\}\,,
\label{phiVy1PN}\end{equation}
where $\zeta=-\frac{743}{672}-\frac{11}{8}\nu$,
$\tau=\frac{3715}{2016}+\frac{55}{24}\nu$. We insert those expressions
into our basic integral (\ref{intK}), and split it into recent past
and remote past contributions. Exactly like in (\ref{remote}) we can
prove that the remote past of that integral, for which $y\in
\bigr]1,+\infty\bigl[$, is negligible --- of the order of ${\cal
O}(\xi^2)$. Next, in the recent-past integral, $y\in \bigr]0,1\bigl[$,
we expand at the 1PN order [{\it i.e.} $x(T_R)\rightarrow 0$], to
obtain some 1PN correction term, with respect to the previous
calculation, of the form
\begin{equation}
\int_0^1\frac{dy\,\ln y}{(1+y)^{5/4}}\,\Bigl[(1+y)^\alpha-1\Bigr]\,
e^{-\frac{32i}{5\xi}\left[(1+y)^{5/8}-1\right]}\,,
\label{int1PN}\end{equation}
where $\alpha$ can take the values $-\frac{1}{4},\,\frac{3}{8}$ or
$\frac{5}{8}$. Now the point is that this integral, like the
remote-past one, is {\it also} of the order of ${\cal O}(\xi^2)$ or,
rather, ${\cal O}(\xi^2\ln\xi)$. Indeed, the new factor
$(1+y)^\alpha-1$ in the integrand of (\ref{int1PN}) is crucial in that
it adds (after taking the equivalent when $y\rightarrow 0$) an extra
factor $y$, and we have thus to treat the following equivalent,
\begin{equation}
\int_0^1 dy\,y\,\ln y \, e^{-\frac{4i\,y}{\xi}} = \mathcal{O}
\left(\xi^2\ln\xi\right)\,,
\label{recent1PN}\end{equation}
which is {\it smaller} by a factor $\xi={\cal O}(5)$ than the integral
(\ref{recent'}), as easily seen by integrating by parts. This means
that the order of magnitude of the correction induced by our more
sophisticated 1PN model for inspiral in the past is negligible. In
conclusion, even at the relative PN order, one can use
Eq.~(\ref{intKres}) for computing the tails, but of course the current
values of the binary's orbital parameters $x(T_R)$, $\phi (T_R)$ and
$\xi(T_R)$ must consistently include their relevant PN
corrections. This is what we do in the present paper, following the
computation in Refs. \cite{Beflux-96,Btail} of the higher-order tails
up to relative 2PN order ({\it i.e.} 3.5PN beyond quadrupolar
radiation).

Finally our results for the 2.5PN-accurate tail terms are as
follows. It is convenient, following \cite{BIWW}, to introduce, in
place of the ``natural'' constant time-scale $b$ entering the tails
and defined by (\ref{b}), a constant {\it frequency}-scale $\omega_0$
given by
\begin{equation}
\omega_0 \equiv \frac{e^{\frac{11}{12}-C}}{4\,b}\,.
\label{om0}\end{equation}
Like $b$, it can be chosen at will, for instance to be equal, as
suggested in \cite{BIWW}, to the seismic cut-off frequency of some
interferometric detector, $\omega_0=\omega_{\rm seismic}$. Then we
find that all the dependence of the tails in the {\it logarithm} of
the orbital frequency, {\it i.e.} the terms involving $\ln\omega$ and
coming from the logarithm present in the R.H.S. of (\ref{recent'}),
can be factorized out, up to the 2.5PN order, in the way
\begin{equation}
\bigl(h_{+,\times} \bigr)_{{\rm tail}} = \bigl(k_{+,\times}
\bigr)_{{\rm tail}}
-2\,x^{3/2}\,\left[1-\frac{\nu}{2}x\right]\frac{\partial
h_{+,\times}}{\partial\phi}\,\ln\left(\frac{\omega}{
\omega_0}\right)\,,
\label{factorlog}\end{equation}
where all the dependence upon $\ln (\omega/\omega_0)$ is given as
indicated [{\it i.e.} the $(k_{+,\times})_{{\rm tail}}$'s are
independent of $\ln (\omega/\omega_0)$]. In the above expression the
factor $1-\frac{\nu}{2}x$ comes from the relation between the total
ADM mass $M$ and the binary's rest mass $m=m_1+m_2$,
namely $M=m\left[1-\frac{\nu}{2}x\right]$. Because we are computing the tails
with 1PN relative precision, this means that the factor of $\ln
(\omega/\omega_0)$, namely $\partial h_{+,\times}/\partial\phi$ and
therefore also $h_{+,\times}$ itself, is given at the relative 1PN
order. The existence of this structure implies an elegant formulation
of the 2.5PN waveform in terms of a new phase variable $\psi$ given by
Eq.~(\ref{psi}) below. The phase $\psi$ was already introduced in
\cite{BIWW}, and we have shown here that it is also valid,
interestingly enough, for tails at the relative 1PN order. The
``main'' tails contributions are then given, up to 2.5PN order, by
{\allowdisplaybreaks
\begin{subequations}\label{ktails}
\begin{eqnarray}
\bigl(k_{+} \bigr)_{{\rm tail}} &=& \frac{2\,G \,m \,\nu \,x}{c^2 R}
\left\{ -2\pi x^{3/2} (1+c_i^2)\cos 2\phi
\right.\nonumber\\&&\qquad\left.+ \frac{s_i}{40} \frac{\delta m}{m}
x^2 \left[\left(11+7c_i^2+10(5+c_i^2)\ln 2\right)\sin \phi
\right.\right.\nonumber\\&&\qquad\qquad\left.\left. -5\pi
(5+c_i^2)\cos \phi -27[7-10\ln (3/2)](1+c_i^2)\sin
3\phi\right.\right.\nonumber\\&&\qquad\qquad\left.\left. +135\pi
(1+c_i^2) \cos 3\phi\right] \right.\nonumber\\&&\qquad\left. +
x^{5/2}\left[
\frac{\pi}{3}\left(19+9c_i^2-2c_i^4+\nu(-16+14c_i^2+6c_i^4)\right)\cos
2\phi \right.\right.\nonumber\\&&\qquad\qquad\left.\left. +
\frac{1}{5}\left(-9+14c_i^2+7c_i^4+\nu(27-42c_i^2- 21c_i^4)\right)\sin
2\phi\right.\right.\nonumber\\&&\qquad\qquad\left.\left. -
\frac{16\pi}{3}(1-c_i^4)\left(1-3\nu\right)\cos 4\phi
\right.\right.\nonumber\\&&\qquad\qquad\left.\left. +
\frac{8}{15}(1-c_i^4)(1-3\nu)\left(21-20\ln 2\right)\sin 4\phi\right]
\right\}\,,\\ 
\bigl(k_{\times} \bigr)_{{\rm tail}} &=& \frac{2\,G \,m
\,\nu \,x}{c^2 R} \left\{ -4\pi x^{3/2} c_i \sin 2\phi
\right.\nonumber\\&&\qquad\left.- \frac{3\,s_i\,c_i}{20} \frac{\delta
m}{m} x^2 \left[\left(3+10\ln 2\right)\cos \phi+5\pi\sin \phi
\right.\right.\nonumber\\&&\qquad\qquad\left.\left. -9[7-10\ln
(3/2)]\cos 3\phi -45\pi \sin 3\phi\right]
\right.\nonumber\\&&\qquad\left. + x^{5/2}\left[
\frac{2\pi}{3}c_i\left(13+4\,s_i^2+\nu(2-12\,s_i^2)\right)\sin 2\phi
\right.\right.\nonumber\\&&\qquad\qquad\left.\left. +
\frac{2}{5}c_i(1-3\nu)\left(-6+11\,s_i^2\right)\cos
2\phi\right.\right.\nonumber\\&&\qquad\qquad\left.\left. -
\frac{32\pi}{3}c_i\,s_i^2(1-3\nu) \sin 4\phi
\right.\right.\nonumber\\&&\qquad\qquad\left.\left. +
\frac{16}{15}c_i\,s_i^2(1-3\nu)\left(-21+20\ln 2\right)\cos 4\phi\right]
\right\}\,.
\end{eqnarray}\end{subequations}}\noindent
Here $c_i$ and $s_i$ denote the cosine and sine of the inclination
angle $i$, and $\delta m = m_1-m_2$ is the mass difference (so that
$\delta m/m = X_1-X_2$). Up to the 2PN order, we have agreement with
the results of \cite{BIWW}.

\section{Results for the 2.5PN polarisation waveforms}\label{sec:2.5PN pol}

\subsection{Instantaneous terms at the 2.5PN order}
\label{sec:2.5PN inst}

In Section \ref{sec:multipole moments} we have given the list of
source multipole moments needed to control the waveform at the 2.5PN
order. In this Section we proceed to calculate the instantaneous terms
[of types (s), (r) and (c)] in the 2.5PN waveform of circular compact
binaries. The first step towards it is the computation of time
derivatives of different moments $I_L$, $J_L$ (and also $W$) using the
binary's EOM up to 2.5PN order as given in Eqs.~(\ref{eq:EOM
2.5PN})--(\ref{eq:gamma-x}). These time derivatives are then contracted
with the unit direction ${\bf N}$, for insertion into
Eqs.~(\ref{eq:WFinst}), (\ref{eq:WFinst(r)}) and
(\ref{eq:WFinst(c)}). We can write down the inst(s) waveform
schematically as,
\begin{eqnarray}\label{eq:WF-schematic}
\bigl(h^{\rm TT}_{km}\bigr)_{\rm inst(s)} &=& \frac{2G\,\nu\,
m}{c^4R}\, {\cal P}_{ijkm}\left\{ \xi^{(0)}_{ij}+ (X_2-X_1)
\xi^{(1/2)}_{ij}+\xi^{(1)}_{ij}\right.\nonumber\\ &&\qquad +
(X_2-X_1)\, \xi^{(3/2)}_{ij}
+\xi^{(2)}_{ij}+\left.(X_2-X_1)\,\xi^{(5/2)}_{ij}+\rho_{ij}^{(5/2)}\right\}\,.
\end{eqnarray}
The instantaneous terms up to 2PN order were already reported in
Eqs. (4.5) of Ref. \cite{BDI}. We have reproduced them in our present
computation. At the 2.5PN order, most of the terms vanish when the two
masses are equal ($X_1=X_2$) like at the previous ``odd
approximations'' 0.5PN and 1.5PN, but there is also an extra
contribution, denoted by $\rho_{ij}^{(5/2)}$ in
Eq.~(\ref{eq:WF-schematic}). This consists of two parts:
$\bigl(\rho_{ij}^{(5/2)}\bigr)_{\rm
reac}=-{64\over5}\,\nu\,\frac{Gm}{r}\frac{\gamma^2}{c}n^{(i}v^{j)}$
which arises directly from the 2.5PN radiation-reaction term in the
EOM given by Eq.~(\ref{eq:EOM 2.5PN}); and
$\bigl(\rho_{ij}^{(5/2)}\bigr)_{\rm
quad}=-{192\over7}\,\nu\,\frac{Gm}{r}\frac{\gamma^2}{c}n^{(i}v^{j)}$
which comes from the 2.5PN contribution in the mass quadrupole
(\ref{eq:massmom}). We find
\begin{equation}
\rho_{ij}^{(5/2)}=-{1408\over35}\,\nu\,\frac{Gm}{r}\frac{\gamma^2}{c}n^{(i}v^{j)}\,.
\end{equation}
The other contributions follow from a long but
straightforward computation starting from the multipole moments listed
earlier, and  read as 
{\allowdisplaybreaks
\begin{eqnarray}\label{eq:5halfinst}
\xi^{(5/2)}_{ij}&=&
\Biggl\{n^{ij}\left[\frac{\gamma}{c^3}\left\{\left({1\over3} -
{4\over3}\,\nu+\,\nu^2\right)({\bf N .v})^5\right\}\right. \\
\nonumber &&~~+\frac{\gamma^2}{c}\left\{\left({1199\over180} -
{539\over45}\,\nu -{101\over60}\,\nu^2\right)({\bf N .v})^3\right. \\
\nonumber &&~~\qquad\left.+\left({263\over10} -{526\over 5}\,\nu
+{789\over10}\,\nu^2\right)({\bf N . n})^2({\bf N .v})^3\right\} \\
\nonumber
&&~~+\frac{Gm}{r}\frac{\gamma^2}{c}\left\{\left(-{263\over72} +
{553\over90}\,\nu+{17\over120}\,\nu^2\right)({\bf N .v})\right. \\
\nonumber &&~~\qquad\left.+\left(-{757\over12}+{8237\over60}\,\nu
-{433\over20}\,\nu^2\right)({\bf N . n})^2({\bf N .v}) \right. \\
\nonumber &&~~\qquad\left.\left. +\left(-{ 2341\over72} +
{2341\over18}\,\nu - {2341\over24}\,\nu^2\right)({\bf N . n})^4({\bf N
.v})\right\}\right] \\ \nonumber &&
+n^{(i}v^{j)}\left[\frac{\gamma}{c^3}\left\{\left (-{74\over3} +
{296\over3}\,\nu -{74}\,\nu^2\right)({\bf N . n})({\bf N .v})^4\right
\}\right.\\ \nonumber &&~~+\frac{\gamma^2}{c}\left\{\left({5161\over90}
-{5612\over45}\,\nu+{461\over30}\,\nu^2\right)({\bf N . n})({\bf N
.v})^2\right.\\ \nonumber &&~~\qquad\left. +\left({1811\over15}
-{7244\over15}\,\nu + {1811\over5}\,\nu^2\right)({\bf N . n})^3({\bf N
.v})^2 \right\} \\ \nonumber
&&~~+\frac{Gm}{r}\frac{\gamma^2}{c}\left\{\left(-{479\over60}
+{187\over6}\,\nu - {9\over4}\,\nu^2\right)({\bf N . n})\right. \\
\nonumber &&~~\qquad\left. +\left(-
{5587\over90}+{1282\over9}\,\nu-{65\over2}\,\nu^2\right)({\bf N
. n})^3\right. \\ \nonumber &&~~\qquad\left.\left. +\left(-
{3787\over180} + {3787\over45}\,\nu - {3787\over60}\,\nu^2\right)({\bf
N . n})^5\right \} \right] \\ \nonumber
&&+v^{ij}\left[\frac{1}{c^5}\left(2 - 8\,\nu + 6\,\nu^2\right)({\bf N
.v})^5 \right.\\ \nonumber &&~~+
\frac{\gamma}{c^3}\left\{\left(-{4\over3} +
{10\over3}\,\nu\right)({\bf N .v})^3 +\left(-{158\over3} +
{632\over3}\,\nu - 158\,\nu^2\right)({\bf N . n})^2({\bf N
.v})^3\right\} \\ \nonumber &&~~
+\frac{\gamma^2}{c}\left\{\left(-{536\over45} -{1531\over45}\,\nu -
{1\over15}\,\nu^2\right)({\bf N .v})\right. \\ \nonumber
&&\qquad~~\left.+\left({1345\over36} - {799 \over9}\,\nu +
{245\over12}\,\nu^2\right)({\bf N . n})^2({\bf N .v})\right. \\
\nonumber &&\qquad~~\left.\left.+\left({2833 \over60} -
{2833\over15}\,\nu + {2833\over20}\,\nu^2\right)({\bf N . n})^4{\bf N
.v}) \right\}\right]\Biggr\}\,,
\end{eqnarray}}\noindent
where we recall that ${\bf n}={\bf x}/r$ and the parameter $\gamma$ is
defined by (\ref{gam}).

In addition we must compute the instantaneous (r) and (c) parts of the
waveform for the compact binaries in circular orbits. These parts are
purely of order 2.5PN. The inst(c) part is computed starting from the
expression for $W$ in Eq.~(\ref{monopoleW}), but it turns out to be
zero for circular orbits. We find
\begin{subequations}\label{eq:5halfinst(rc)}\begin{eqnarray}
\bigl(h^{\rm TT}_{km}\bigr)_{\rm inst(r)} &=& \frac{2G\,\nu\,
m}{c^4R}\, {\cal P}_{ijkm} {\nu\,\gamma^2 \over c} \biggl\{ -{84 \over
5}\frac{G \,m}{r}({\bf N . n})({\bf N . v})\,n^{ij} +28 ({\bf N
. n})({\bf N . v})v^{ij}\nonumber\\ &&\qquad +\frac{G \,m}{r}\left({64
\over35}-{192 \over5}({\bf N . n})^2\right)n^{(i}v^{j)} +16 ({\bf
N. v})^2 n^{(i}v^{j)} \biggr\}\,,\\ \bigl(h^{\rm TT}_{km}\bigr)_{\rm
inst(c)} &=& 0\,.
\end{eqnarray}\end{subequations}

\subsection{The complete plus and cross polarisations}\label{complete}

We have computed all the five different contributions to the waveform
contained in Eq.~(\ref{eq:wf-parts}). This together with the results
of \cite{BDI} provides a complete 2.5PN accurate waveform for the
circular orbit case. In this Section, from the 2.5PN waveform, we
present the result for the two gravitational-wave polarisations,
extending a similar analysis at the 2PN order in Ref. \cite{BIWW}.

The polarisations corresponding to the instantaneous terms are
computed using Eqs.~(\ref{eq:5halfinst}) and (\ref{eq:5halfinst(rc)}),
while those corresponding to the hereditary terms where obtained in
Eqs.~(\ref{plus-crossresmem}) and (\ref{factorlog})--(\ref{ktails}). As
in the earlier work, these polarisations are represented in terms of
the gauge invariant parameter $x\equiv({Gm\omega /c^3})^{2/3}$, where
$\omega$ represents the orbital frequency of the circular orbit,
accurate up to 2.5PN order. This requires the relation between
$\gamma$ and $x$, which has already been given in
Eq.~(\ref{eq:gamma-x}). The final form of the 2.5PN polarisations may
now be written as,
\begin{equation}\label{eq:pol-schematic}
h_{+,\times} = \frac{2\,G \,m \,\nu
\,x}{c^2\,R}\left\{H_{+,\times}^{(0)}+x^{1/2}\,H_{+,\times}^{(1/2)}
+x\,H_{+,\times}^{(1)}+x^{3/2} \,H_{+,\times}^{(3/2)}
+x^2\,H_{+,\times}^{(2)}+x^{5/2}\,H_{+,\times}^{(5/2)}\right\}\,.
\end{equation}
In particular we shall recover the 2PN results of
\cite{BIWW}. However, for the comparison we have to employ the same
phase variable as in \cite{BIWW}, which means introducing an auxiliary
phase variable $\psi$, shifted away from the actual orbital phase
$\phi$ we have used up to now, by Eq. (5) of \cite{BIWW}. Furthermore,
the phase $\psi$ given in \cite{BIWW} is {\it a priori} adequate up to
only the 2PN order, but we have proved it to be also correct at the
higher 2.5PN order with the mass in front of the log-term being the
ADM mass. Indeed, the motivation for the shift
$\phi\longrightarrow\psi$ is to ``remove'' all the logarithms of the
frequency [{\it i.e.} $\ln\omega$ or, rather, $\ln (\omega/\omega_0)$]
in the two polarisation waveforms and to absorb them into the
definition of the new phase angle $\psi$. As a result, the two
polarisation waveforms, when expressed in terms of $\psi$ instead of
$\phi$, become substantially simpler. From Eq.~(\ref{factorlog}) we
see that if we re-express the waveform by means of the phase
\cite{BIWW}
\begin{equation}
\psi=\phi- 2
x^{3/2}\left[1-\frac{\nu}{2}x\right]\ln\left({\omega\over\omega_0}\right)\,,
\label{psi}\end{equation}
we are able to move all the $\ln\omega$-terms into the phase
angle. Notice that the possibility of this move is interesting because
it shows that in fact the $\ln\omega$-terms, which were originally
computed as some modification of the wave {\it amplitude} at orders
1.5PN, 2PN and 2.5PN, now appear as a modulation of the {\it phase} of
the wave at the relative orders 4PN, 4.5PN and 5PN. The reason is that
the lowest-order phase evolution is at the inverse of the order of
radiation-reaction, {\it i.e.} $c^5={\cal O}(-5)$, so as usual there
is a difference of 2.5PN order between amplitude and phase. This shows
therefore that the modification of the phase in (\ref{psi}) is
presently negligible (it is of the same order of magnitude as unknown
4PN terms in the orbital phase evolution when it is given as a
function of time). It could be ignored in practice, but it is probably
better to keep it as it stands into the definition of templates of
ICBs. The phase shift (\ref{psi}) corresponds to some spreading of the
different frequency components of the wave, {\it i.e.} the ``wave
packets'' composing it, along the line of sight from the source to the
detector (see \cite{BS-93} for a discussion).

With this above choice of phase variable, the same as in \cite{BIWW},
all terms up to 2PN match with those listed in Eqs. (3) and (4) of
\cite{BIWW}, though we recast them in a slightly different form for
our convenience to present the 2.5PN terms. We find,
\allowdisplaybreaks{
\begin{subequations}\label{eq:biww-plus}
\begin{eqnarray}
H^{(0)}_+&=&-(1+\,c_i^2) \cos
2\psi-{1\over96}\,\,s_i^2\,(17+\,c_i^2)\,,\label{H0memory} \\
H^{(0.5)}_+&=&-\,s_i {{\delta m \over m}}\left[\cos\psi
\left({5\over8}+{1\over8}\,c_i^2\right)-\cos
3\psi\left({9\over8}+{9\over8}\,c_i^2\right)\right]\,, \\
H^{(1)}_+&=&\cos2\psi\left[{19\over6} + {3\over2}\,c_i^2-{
1\over3}\,c_i^4 +\nu\left(-{19\over6}+{11\over6}\,c_i^2
+\,c_i^4\right)\right]\nonumber\\
&&-\cos4\psi\left[{4\over3}\,s_i^2(1+\,c_i^2)(1-3\nu)\right]\,,\\
H^{(1.5)}_+ &=&\,s_i{{\delta m \over m}}\cos\psi\left[{19\over64}
+{5\over16}\,c_i^2-{1\over192}\,c_i^4 +\nu\left(-{49\over96}
+{1\over8}\,c_i^2+{1\over96}\,c_i^4\right)\right]\nonumber\\
&&+\cos2\psi\left[-2\pi(1+\,c_i^2)\right]\nonumber\\ &&+\,s_i{{\delta
m \over m}}\cos3\psi\left[ -{657\over128}
-{45\over16}\,c_i^2+{81\over128}\,c_i^4 \right.\nonumber\\
&&~~+\left.\nu\left({225\over64}-{9\over8}\,c_i^2
-{81\over64}\,c_i^4\right)\right]\nonumber\\ &&+\,s_i{{\delta m \over
m}}\cos5\psi\left[{625\over 384}\,s_i^2(1+\,c_i^2)(1
-2\nu)\right]\,,\\ H^{(2)}_+ &=&\,\pi \,s_i\,{\delta m \over
m}\cos\psi\left[-{5\over8} -{1\over8}\,c_i^2\right]\nonumber\\
&&+\cos2\psi\left[ {11\over60}+{33\over10}\,c_i^2+{29\over24}\,c_i^4
-{1\over24}\,c_i^6 \right.\nonumber\\
&&~~+\nu\left({353\over36}-3\,c_i^2-{251\over72}\,c_i^4
+{5\over24}\,c_i^6\right)\nonumber\\
&&~~+\left.\nu^2\left(-{49\over12}+{9\over2}\,c_i^2-{7\over24}\,c_i^4
-{5\over24}\,c_i^6\right)\right]\nonumber\\ &&+\pi \,s_i{{\delta m
\over m}}\cos3\psi\left[{27\over8}(1+\,c_i^2)\right]\nonumber\\
&&+\cos4\psi\left[ {118\over15}-{16\over5}\,c_i^2-{86\over15}\,c_i^4
+{16\over15}\,c_i^6 \right.\nonumber\\
&&~~+\left.\nu\left(-{262\over9}+16\,c_i^2+{166\over9}\,c_i^4
-{16\over3}\,c_i^6\right)\right.\nonumber\\
&&~~+\left.\nu^2\left(14-16\,c_i^2-{10\over3}\,c_i^4
+{16\over3}\,c_i^6\right)\right]\nonumber\\
&&+\cos6\psi\left[-{81\over40}\,s_i^4(1+\,c_i^2)\left(1-5\nu
+5\nu^2\right)\right]\nonumber\\ &&+\,s_i{\delta m \over m}
\sin\psi\left[{11\over40}+{5\ln2\over
4}+\,c_i^2\left({7\over40}+{\ln2\over4}\right)\right]\nonumber\\
&&+\,s_i{\delta m \over m} \sin3\psi\left[\left(-{189\over
40}+{27\over4}\ln(3/2)\right)(1+\,c_i^2)\right]\,,
\end{eqnarray}
\end{subequations}}
\allowdisplaybreaks{\begin{subequations}\label{eq:biww-cross}
\begin{eqnarray}
H^{(0)}_\times&=&-2c_i \sin2\psi\,,\\ H^{(0.5)}_\times&=&
 \,s_ic_i\,{\delta m \over
 m}\left[-{3\over4}\sin\psi+{9\over4}\sin3\psi\right]\,,\\
 H^{(1)}_\times&=& c_i \sin2\psi\left[{17\over3}-{4\over3}\,c_i^2
 +\nu\left(-{13\over3}+4\,c_i^2\right)\right]\nonumber\\&&+c_i
 \,s_i^2\sin4\psi\left[-{8\over3}(1-3\nu)\right]\,,\\
 H^{(1.5)}_\times&=&\,s_ic_i {\delta m \over m}
 \sin\psi\left[{21\over32} -{5\over96}\,c_i^2 +\nu\left(-{23\over48}
 +{5\over48}\,c_i^2\right)\right]\nonumber\\ &&-4\pi\,c_i
 \sin2\psi\nonumber\\ &&+\,s_ic_i\,{\delta m \over
 m}\,\sin3\psi\left[-{603\over64} +{135\over64}\,c_i^2
 +\nu\left({171\over32} -{135\over32}\,c_i^2\right)\right]\nonumber\\
 &&+\,s_ic_i\,{\delta m \over
 m}\,\sin5\psi\left[{625\over192}(1-2\nu)\,s_i^2\right]\,,\\
 H^{(2)}_\times&=&\,s_ic_i\,{\delta m \over
 m}\cos\psi\left[-{9\over20}
 -{3\over2}\ln2\right]\nonumber\\&&+\,s_ic_i\,{\delta m \over
 m}\cos3\psi\left[{189\over20}-{27\over2}\ln(3/2)\right]\nonumber\\
 &&-\,s_ic_i\,{\delta m \over m}\left[{3\,\pi\over4}\right]
 \sin\psi\nonumber\\
 &&+c_i\sin2\psi\left[{17\over15}+{113\over30}\,c_i^2
 -{1\over4}\,c_i^4 \right.\nonumber\\
 &&~~+\nu\left({143\over9}-{245\over18}\,c_i^2+{5\over4}\,c_i^4\right)\nonumber\\
 &&~~+\left.\nu^2\left(-{14\over3}+{35\over6}\,c_i^2
 -{5\over4}c_i^4\right)\right]\nonumber\\ &&+\,s_ic_i\,{\delta m \over
 m}\sin3\psi\left[{27\pi\over4}\right]\nonumber\\
 &&+c_i\sin4\psi\left[{44\over3}-{268\over15}\,c_i^2
 +{16\over5}\,c_i^4 \right.\nonumber\\
 &&~~+\nu\left(-{476\over9}+{620\over9}\,c_i^2-16\,c_i^4\right)\nonumber\\
 &&~~+\left.\nu^2\left({68\over3}-{116\over3}\,c_i^2
 +16\,c_i^4\right)\right]\nonumber\\
 &&+c_i\sin6\psi\left[-{81\over20}\,s_i^4(1-5\nu+5\nu^2)\right]\,.
\end{eqnarray}
\end{subequations}}\noindent
Notice a difference with the results of \cite{BIWW}, in that we have
included the specific effect of non-linear memory into the
polarization waveform at the {\it Newtonian} order, {\it c.f.} the
term proportional to $s_i^2(17+c_i^2)$ in $H^{(0)}_+$ given by
(\ref{H0memory}) above. This is consistent with the order of magnitude
of this effect, calculated in Section \ref{sec:NLM}. However, beware
of the fact that the memory effect is rather sensitive on the details
of the whole time-evolution of the binary prior the current detection,
so the zero-frequency (DC) term we have included into (\ref{H0memory})
may change depending on the binary's past history (see our discussion
in Section \ref{sec:NLM}). Nevertheless, we feel that it is a good
point to include the ``Newtonian'' non-linear memory effect, exactly
as it is given in Eq.~(\ref{H0memory}), for the detection and analysis
of ICBs.\footnote{We already remarked that we have not computed the DC
terms possibly present in the higher-order harmonics of the 2.5PN
waveform.}

The purely 2.5PN contributions, in the plus and cross polarisations,
constitute, together with the memory term in (\ref{H0memory}), the
final result of this paper. They read as, {\allowdisplaybreaks
\begin{eqnarray}\label{H+final}
H^{(2.5)}_+ &=& s_i{\delta m\over m}\cos\psi\left[{1771\over
5120}-{1667\over5120}\,c_i^2+{217\over 9216}\,c_i^4-{1\over
9216}\,c_i^6\right.\\ \nonumber
&&~~+\nu\left({681\over256}+{13\over768}\,c_i^2-{35\over
768}\,c_i^4+{1\over2304}\,c_i^6\right)\\ \nonumber
&&~~+\left.\nu^2\left(-{3451\over9216}+{673\over3072}\,c_i^2-
{5\over9216}\,c_i^4-{1\over3072}\,c_i^6\right)\right]\\ \nonumber
&&+{\pi}\cos2\psi\left[{19\over3}+3\,c_i^2-{2\over3}\,c_i^4+\nu
\left(-{16\over3}+{14\over3}\,c_i^2+2\,c_i^4\right)\right]\\ \nonumber
&&+s_i{\delta m\over
m}\cos3\psi\left[{3537\over1024}-{22977\over5120}\,c_i^2-{15309
\over5120}\,c_i^4+{729\over5120}\,c_i^6\right.\\ \nonumber
&&~~+\nu\left(-{23829\over1280}+{5529\over1280}\,c_i^2+{7749\over1280}
\,c_i^4-{729\over1280}\,c_i^6\right)\\ \nonumber
&&~~+\left.\nu^2\left({29127\over5120}-{27267\over5120}\,c_i^2-{1647
\over5120}\,c_i^4+{2187\over 5120}\,c_i^6\right)\right]\\ \nonumber
&&+\cos4\psi\left[-{16\pi\over3}\,(1+\,c_i^2)\,s_i^2(1-3\nu)\right]\\
\nonumber &&+s_i {\delta m\over
m}\cos5\psi\left[-{108125\over9216}+{40625\over9216}\,c_i^2
+{83125\over9216}\,c_i^4-{15625\over 9216}\,c_i^6\right.\\ \nonumber
&&~~+\nu\left({8125\over
256}-{40625\over2304}\,c_i^2-{48125\over2304}\,c_i^4+{15625\over
2304}\,c_i^6\right)\\ \nonumber
&&~~+\left.\nu^2\left(-{119375\over9216}+{40625\over
3072}\,c_i^2+{44375\over 9216}\,c_i^4-{15625\over
3072}\,c_i^6\right)\right]\\ \nonumber &&+{\delta m\over m}
\cos7\psi\left[{117649\over46080}\,s_i^5(1+\,c_i^2)(1-4\nu+3\nu^2)
\right]\\\nonumber
&&+\sin2\psi\left[-{9\over5}+{14\over5}\,c_i^2+{7\over5}\,c_i^4+\nu
\left({96\over5}-{8\over5}\,c_i^2-{28\over5}\,c_i^4\right)\right]\\
\nonumber
&&+s_i^2(1+\,c_i^2)\sin4\psi\left[{56\over5}-{32\ln2\,\over3}
-\nu\left({1193\over30}-32\ln2 \right)\right]\,.
\end{eqnarray}}
\allowdisplaybreaks{\begin{eqnarray}\label{Hxfinal} H^{(2.5)}_\times
&=&{6\over5}\,s_i^2\,c_i\,\nu\\\nonumber
&&+\,c_i\cos2\psi\left[2-{22\over5}\,c_i^2+\nu\left(-{154\over5}+{94\over5}
\,c_i^2\right)\right]\\ \nonumber &&+\,c_i\,s_i^2
\cos4\psi\left[-{112\over5}+{64\over3}\ln2+\nu\left({1193\over15}-64\ln2
\right)\right]\\\nonumber &&+\,s_i\,c_i{\delta m \over m}
\sin\psi\left[-{913\over 7680}+{1891\over11520}
\,c_i^2-{7\over4608}\,c_i^4\right.\\\nonumber
&&~~+\left.\nu\left({1165\over384}-{235\over576}\,c_i^2
+{7\over1152}\,c_i^4\right)\right.\\\nonumber
&&~~+\left.\nu^2\left(-{1301\over4608}+{301\over
2304}\,c_i^2-{7\over1536}\,c_i^4\right)\right]\\\nonumber &&+\pi
\,c_i\sin2\psi\left[{34\over3}-{8\over3}\,c_i^2-\nu\left({20\over3}
-8\,c_i^2\right)\right]\\\nonumber &&+\,s_i\,c_i{\delta m \over
m}\sin3\psi\left[{12501\over2560}-{12069\over1280}\,c_i^2+{1701\over2560}
\,c_i^4\right.\\\nonumber
&&~~+\nu\left(-{19581\over640}+{7821\over320}\,c_i^2-{1701\over640}
\,c_i^4\right)\\\nonumber &&~~+\left.\nu^2\left({18903\over
2560}-{11403\over
1280}\,c_i^2+{5103\over2560}\,c_i^4\right)\right]\\\nonumber
&&+\,s_i^2\,c_i\sin4\psi\left[-{32\pi
\over3}(1-3\nu)\right]\\\nonumber &&+{\delta m \over m}
\,s_i\,c_i\sin5\psi\left[-{101875\over4608}+{6875\over256}\,c_i^2-
{21875\over4608}\,c_i^4\right.\\\nonumber
&&~~+\left.\nu\left({66875\over1152}-{44375\over576}\,c_i^2+{21875\over1152}
\,c_i^4\right)\right.\\\nonumber
&&~~+\left.\nu^2\left(-{100625\over4608}+{83125\over2304}\,c_i^2-{21875\over1536}
\,c_i^4\right)\right]\\\nonumber &&+{\delta m\over
m}\,s_i^5\,c_i\sin7\psi\left[{117649\over23040}\left(1-4\nu+3\nu^2\right)\right]
\,.
\end{eqnarray}}\noindent
Note that the latter cross polarisation contains a zero-frequency term
[first term in Eq. (\ref{Hxfinal})], which comes from the inst(r)
contribution given by (\ref{eq:5halfinst(rc)}). We employ the same
notation as in \cite{BIWW}, except that $c_i$ and $s_i$ denote
respectively cosine and sine of the inclination angle $i$ (which is
defined as the angle between the vector ${\bf N}$, along the line of
sight from the binary to the detector, and the normal to the orbital
plane, chosen to be right handed with respect to the sense of motion,
so that $0 \leq i \leq \pi$). In particular the mass difference reads
$\delta m = m_1-m_2$. Like in \cite{BIWW}, our results are in terms of
the phase variable $\psi$ defined by (\ref{psi}) in function of the
actual orbital phase $\phi$ (namely the angle oriented in the sense of
motion between the ascending node ${\cal N}$ and direction of body one
--- {\it i.e.} $\phi=0$ mod $2\pi$ when the two bodies lie along ${\bf
P}$). We have verified that the plus and cross polarizations (\ref{H+final})-(\ref{Hxfinal}) reduce in the limit $\nu\rightarrow 0$ to the result of black
hole perturbation theory as given in the Appendix B of Tagoshi and Sasaki 
\cite{TS94} (the phase variable used in \cite{TS94} differs from ours by
$\psi_{\rm TS}=\psi+\pi/2+2x^{3/2}[\ln 2-17/12]$ and we have
  $\theta_{\rm TS}=\pi- i$).\footnote{We spotted a misprint in the Appendix B of \cite{TS94}, namely the sign of the harmonic coefficient $\zeta_{7,3}^\times$ (\textit{i.e.} having $l=7$, $m=3$, and corresponding to the cross polarisation) should be changed, so that one should read $\zeta_{7,3}^\times=+
 \frac{729}{10250240}\cos(\theta)(167+...)\sin(\theta)(v^5 \cos(3 \psi)-...)$.}.

Equations (\ref{H+final}) and (\ref{Hxfinal}), together with
(\ref{eq:biww-plus})--(\ref{eq:biww-cross}), provide the 2.5PN
accurate template for the ICBs moving on quasi-circular orbits,
extending the results of \cite{BIWW} by half a PN order. They are
complete except for the possible inclusion of memory-type
(zero-frequency or DC) contributions in higher PN amplitudes (1PN and 2PN). These
wave polarisations together with the phasing formula of
Ref. \cite{BFIJ}, {\it i.e.} the crucial time variation of the phase:
$\phi(t)$, constitute the currently best available templates for the
data analysis of ICB for ground based as well as space-borne GW
interferometers.

\subsection{Comments on the 3PN waveform}\label{further}

In Section \ref{sec:multipole moments} we have given the list of
source multipole moments needed to control the waveform at the 2.5PN
order. The computation of the 3PN waveform obviously requires more
accurate versions of these moments as well as new moments, which all
together would constitute the basis of the computation of the 3PN
waveform. Thus, even though the 3PN mass quadrupole moment \cite{BIJ}
and 3PN accurate EOM
\cite{JS98,JS99,BF00,BFeom,DJS00,DJS01,ABF,BI-03,dimreg,BDE03} are
available, our present level of accuracy is {\it not} sufficient
enough to compute the 3PN waveform.\footnote{We are speaking here of
the 3PN {\it waveform}. The computation of the 3PN {\it flux} is less
demanding, because each multipolar order brings in a new factor
$c^{-2}={\cal O}(2)$ instead of ${\cal O}(1)$ in the case of the
waveform, which explains why it is possible to control it up to the
3.5PN order in \cite{BIJ}.}

The source multipole moments at 3PN order, would yield the control of
the inst(s) part of the waveform, as well as the tail terms, but we
have to consider also other types of contributions, which are not all
under control. The main reason for the present incompleteness at 3PN
order is that the instantaneous terms of type (r) and (c),
generalizing (\ref{eq:WFinst(r)})--(\ref{eq:WFinst(c)}) to the 3PN
order, are not computed.

Recall that the inst(r) terms are those whose sum constitutes the
instantaneous part of the relationships between the radiative moments
$U_L$, $V_L$, and the ``canonical'' ones $M_L$, $S_L$, see
(\ref{eq:U})--(\ref{eq:V}). Though one can guess the structure of
these terms at the 3PN order, using dimensional and parity arguments,
the numerical coefficients in front of each of them require detailed
(and generally tedious) work. For instance, in the 3PN waveform we
shall need the radiative mass-type octupole moment $U_{ijk}$ at the
2.5PN order, and therefore we have to know what is the remainder term
${\cal O}(5)$ in Eq.~(\ref{eq:Uijk}) which we don't (such a
calculation would notably entail controlling the quadratic
interactions between one mass and one current quadrupole,
$M_{ij}\times S_{kl}$, and between one mass quadrupole and one
octupole, $M_{ij}\times M_{klm}$). And similarly for the radiative
current-type quadrupole $V_{ij}$ given by (\ref{eq:Vij}). We have also
to compute the remainder terms ${\cal O}(3)$ in the corresponding
expressions of $U_{ijklm}$ and $V_{ijkl}$.

Concerning the inst(c) terms, which are the instantaneous terms coming
from the difference between the canonical moments $M_L$, $S_L$ and the
general source ones $I_L$, $J_L$, $\cdots$ [{\it c.f.}
Eqs.~(\ref{MijIij})--(\ref{MLIL})], it does not seem to be obvious to
guess even their structure. The crucial new input we would need at 3PN
order concerns the relation between the canonical mass octupole
$M_{ijk}$ and current quadrupole $S_{ij}$ to the corresponding source
moments $I_{ijk}$ and $J_{ij}$ at 2.5PN order, using for instance an
analysis similar to the one in \cite{Beflux-96}.

Finally, at 3PN order we would have to extend the present computation
of hereditary terms. In the case of quadratic tails, like in
(\ref{eq:WFtail}), the computation would probably be straightforward
(indeed we have seen in Section \ref{sec:tail} that the complications
due to the influence of the model of adiabatic inspiral in the past
appear only at the 4PN order), but we have also to take into account
the tail-of-tail cubic contribution in the mass-quadrupole moment at
3PN order, given in Eq. (4.13) of Ref. \cite{Btail}. In addition the
analysis should be extended to the non-linear memory terms. The
complete 3PN waveform and polarisations can be computed only after all
the points listed above are addressed.

\acknowledgments L.B. and B.R.I. thank IFCPAR for its support during
the final stages of this work. L.B. would like to thank the Raman
Research Institute for hospitality while this work was
completed. M.S.S.Q acknowledges Indo-Yemen cultural exchange programme. Almost all algebraic calculations leading to the results of
this paper are done with the aid of the software MATHEMATICA.

\end{document}